\documentclass[final,5p,times,twocolumn]{elsarticle}

\usepackage{latexsym}
\usepackage{amssymb, amsmath, bm, physics}
\usepackage{subcaption}
\usepackage{mathptmx}
\usepackage{flushend}
\usepackage[colorlinks,citecolor=blue,urlcolor=blue,linkcolor=blue]{hyperref}
\usepackage{aas_macros}
\usepackage{orcidlink}

\biboptions{sort&compress}

\begin{document}

\begin{frontmatter}




\title{Probing extra dimensions through cosmological observations of dark energy}

\author{R. Jalalzadeh$^{1}$
\orcidlink{0000-0002-6110-3981}}
\ead{r.jalalzadeh@uok.ac.ir}
\author{S. Jalalzadeh$^{2}$ \orcidlink{0000-0003-4854-2960}}
\ead{shahram.jalalzadeh@ufpe.br}
\author{B. Malekolkalami$^{1}$ \orcidlink{0000-0002-5237-1730}}
\ead{b.malakolkalami@uok.ac.ir}


\address{$^{1}$Department of Physics, University of Kurdistan, Pasdaran St., Sanandaj, Iran}
\address{$^{2}$Departamento de F\'{i}sica, Universidade Federal de Pernambuco,
Recife, PE 50670-901, Brazil}

\begin{abstract}
We investigate the isometrically embedded Bianchi type-V cosmology braneworld model in a $D$-dimensional bulk space. The model provides a fluid of geometric dark energy (GDE) and unification of fundamental forces similar to the Kaluza--Klein (KK) theory. The Planck energy density, the fine structure constant, the muon mass, and the number of extra dimensions are all factors that determine the density of the induced GDE. The model also predicts that graviton has mass, which is determined by the induced cosmological constant (CC). Our results are compatible with observations of the standard model of cosmology if the Universe has 22 non-compact extra dimensions.  Our model provides an alternative method for probing extra dimensions.
\end{abstract}


\end{frontmatter}

\section{Introduction}

There might be more than four dimensions in Universe. This idea captures people's imagination, and it seems as far-fetched now as it did when it was originally put forward at the turn of the 20th century. A higher dimensional space is possible from an experimental standpoint, and tests for its existence are now being conducted. Theoretically, an extended space-time offers answers and new insights into fundamental problems in physics and philosophy. Nordstr\"om \cite{2007p2221N}, Kaluza \cite{Kaluza:1921tu}, and Klein \cite{1926ZPhy95K} are the pioneers of attempts to extend general relativity (GR) to integrate gravity and electromagnetism within a common geometrical framework. The very high dimensionality of superstring theories is their most significant feature, and the field's most pressing challenge is to find a way to condense the $26D$ or $10D$ theory into a physical theory with only four dimensions. In this instance, the dimension of space-time in the Nambu--Goto strings cannot be greater than 26 for a formulation without ghosts. Only $10D$ space-time is found to be anomaly-free in the supersymmetric string theory. Additionally, it is possible to interpret the heterotic string of the Gross, Schwarz, and Green theory as using the trivial identity, $26=16+10$, to compactify the 26 dimensions to 10.

 According to astrophysical observations, just 4\% of the Universe's total energy density is in the form of baryonic matter, 23\% is dark matter (DM), and almost 73\% is of a wholly unknown component with negative pressure. The component with negative pressure is known as dark energy (DE), producing a repulsive force responsible for the Universe's present accelerated expansion.

A CC is the most straightforward approach to explain the Universe's late time acceleration in Robertson--Walker space-time and typically in Bianchi-types. Despite the fact that $\Lambda$ is merely another gravitational constant (along with Newton's gravitational constant, $G_N$), it enters the Einstein equations in the same manner as vacuum energy does, namely via a Lorentz-invariant energy-momentum tensor. Because the sole visible characteristic of the CC and vacuum energy is their influence on space-time, observation cannot differentiate between vacuum energy and a classical CC constant. For a quantum field theory (QFT) with cutoff energy scale $M_\text{c}$,
the vacuum energy density scales with the cutoff as $\rho_\text{(vac)}\simeq M^4_\text{c}$, corresponding to a CC $\Lambda_\text{vac} = 8\pi G_N\rho_\text{(vac)}$. If, for example, we set $E_\text{c}=M_\text{Pl}$ ($M_\text{Pl}=1/\sqrt{G_N}$ is the Planck energy), this yields a naive contribution to the CC of $\Lambda_\text{vac}\simeq 10^{38} \text{GeV}^2$, whereas the measured effective CC is the sum of the ``bare'' CC and the contribution from the cutoff scale $\Lambda_\text{eff}=\Lambda+\Lambda_\text{vac}\simeq 10^{-84} \text{GeV}^2$. Therefore, a cancellation of about 123 orders of magnitude is required. The expectation that quantum gravity would explain this cancellation by showing that the vacuum does not gravitate, $\Lambda_\text{vac}=0$, is a reasonable approach to a complicated issue. In this case, CC is a different gravitational constant unrelated to the vacuum energy \cite{ellismaarte}.

{There are four major approaches (if we assume the space-time is a semi-Riemannian manifold) to investigating the CC problem: 1) modified $4D$ gravity theories, 2) modified matter models, 3) inhomogeneous cosmological models, and 4) multidimensional gravity theories.}

{Models of modified gravity in which the source of DE is characterized as a change in gravity. $f(R)$ gravity \cite{DeFelice:2010aj,Nojiri:2010wj,Clifton:2011jh,Miranda:2009rs,Amendola:2006we}, scalar-tensor theories \cite{Amendola:1999qq,Uzan:1999ch,Bartolo:1999sq,Esposito-Farese:2000pbo}, and Gauss--Bonnet gravity (Gauss--Bonnet term is coupled to
scalar field) \cite{Koivisto:2006xf,Koivisto:2006ai,Tsujikawa:2006ph}, are examples. These ideas modify the rules of gravity to accomplish the Universe's late-time accelerated expansion without the need for an explicit DE component.}

The second alternative is to maintain GR as the theory that underlies gravity while inserting scalar fields or exotic fluids (which typically have descriptions similar to those of scalar fields) into the matter sector of gravity field equations. These dynamical models are referred to as DE when they describe late-time acceleration. {In particle physics, several scalar fields exist, including string theory and supergravity. Quintessence is one of the examples
of modified matter models of DE. The word ``quintessence'' refers to a canonical scalar field $\phi$ with a minimally coupled potential, $V(\phi)$, that only interacts with the other components via classical gravity. The equation of state (EoS) of quintessence, in contrast to the CC, dynamically evolves throughout time. The cosmic dynamics for quintessence in the presence of matter and radiation have a long history; several authors, see for example \cite{Fujii:1982ms,Ford:1987de,Wetterich:1987fm,Ratra:1987rm}, have already studied the cosmological implications for such a system. A dynamical system technique may be used to understand cosmological evolution with ease. Some well-known modified matter models are k-essence \cite{Chiba:1999ka,Armendariz-Picon:2000nqq,Armendariz-Picon:2000ulo}, phantoms \cite{Singh:2003vx,Sami:2003xv,Carroll:2003st}, coupled DE \cite{Amendola:1999er,Dalal:2001dt,Zimdahl:2001ar}, and unified DE and DM \cite{Bertacca:2007ux,Fukuyama:2007sx} models. These models are more intricate, but they also have phenomenologically rich characteristics.}

{Inhomogeneous cosmological models include the void \cite{Tomita:1999qn,Tomita:2000jj,Celerier:1999hp,Alnes:2005rw,19865O,Hashemi:2014lqa} and the backreaction models \cite{Rasanen:2003fy,Kolb:2005da,Hirata:2005ei}. Void models desire to supply supernovae data with apparent acceleration caused by strong inhomogeneity. The backreaction model, on the other hand, attempts to explain cosmic acceleration by arranging inhomogeneities in such a way that the departure from the Friedmann--Lemaître--Robertson--Walker (FLRW) metric may provide a precise acceleration. The backreaction model is based on two assumptions: that the GR equations are nonlinear and that the Universe is not homogenous, at least on small scales and maybe on super-horizon scales. As a result, the traditional strategy of averaging the inhomogeneities and then solving the GR equations may differ from first solving the whole inhomogeneous GR equations and then averaging them.}

The fundamental theory of gravity is completely altered by a fourth method, occasionally motivated by the unification of gravity with other gauge forces.
These theories frequently include extra dimensions, producing a $4D$ theory that works as a low energy limiting case. {These models range from diverse brane models to KK (and its extensions) models. For example, the conventional KK gravity action is introduced in Ref. \cite{Waeming:2021ytf} (in the same direction, see \cite{Pongkitivanichkul:2020txi}) together with an extra scalar field and two gauge fields. Whereas a gauge field from $5D$ metric forms a set of mutually orthogonal vectors with two extra gauge fields, the compactification procedure creates a Brans--Dicke model with a dilaton related to the tower of scalar fields. The introduction of the barotropic material completes the realistic setup. They demonstrate that the model may offer the critical points necessary for identifying the presence of DM, DE, and phantom DE by simply including barotropic matter. The author of Ref. \cite{PoncedeLeon:2010kh} also demonstrates how Brans--Dicke's theory (sometimes called the Jordan--Brans--Dicke theory) in $5D$ may explain the current acceleration of cosmic expansion without referring to matter fields in $5D$ or DE in $4D$. He shows that a Brans--Dicke theory in $4D$ with a self-interacting potential and an effective matter field is identical to the vacuum Brans--Dicke field equations in $5D$. The resulting reduced theory provides models for an accelerated expansion of a matter-dominated Universe in the setting of FLRW cosmologies. Moreover, Ref. \cite{Chen:2009ep} has investigated the KK theory's torsion cosmology model. They conclude that the action of torsion can cause the late time acceleration of the Universe in the absence of DE}
\cite{Pietroni:2002ey}. {Evidently, one of the reasons why large and infinite extra dimensions are intriguing is the possibility of their detection. There are a variety of proposals in the literature to detect extra dimensions, including astrophysical observations,
precision tests of gravity on small scales, and collider
searches \cite{Antoniadis:1998ig,Johannsen:2008aa,Salumbides:2015qwa,Chakravarti:2019aup,Pardo:2018ipy,Vagnozzi:2019apd,Visinelli:2017bny,Corman:2020pyr,Corman:2021avn,Du:2020rlx}.  }

The first models of higher dimensions are those put out by Kaluza \cite{Kaluza:1921tu} and Klein \cite{1926ZPhy95K}. The extra dimensions in KK theories were compactified to prevent discrepancies with our $4D$ world \cite{Rasouli:2022tmc}. Braneworld scenarios are also created by moving in this third direction, but unlike the KK theory, these frameworks no longer call for the compactification of the extra dimensions. Initially phenomenological models, these ideas eventually found new life within the context of string theory.

The fundamental hypothesis is that our world is a submanifold (brane) embedded in a higher dimensional bulk space. All matter fields are confined within the brane, while gravity can propagate through this bulk. DE is usually realized in these models as a bulk-generated effect.the Dvali--Gabadadze--Porrati (DGP) braneworld \cite{Deffayet:2001pu,Deffayet:2002sp,Dvali:2000xg,Deffayet:2000uy}, the Randall and Sundrum (RS II) braneworld \cite{Randall:1999vf}, the P. S. Wesson's Induced Matter Theory \cite{2004CQGra611M,doi:10.1142/6029,doi:10.1142/10871,Jalalzadeh:2006mr,Jalalzadeh:2008xu},  and the Covariant Extrinsic Gravity of M. D. Maia \cite{1986GReGr695M,1994PhRvD7233M,2002IJMPA341M,2002PhLA9M,2011GReGr2685M,Jalalzadeh:2013wza,Rostami:2015ixa,Heydari-Fard:2006klr} are all examples of brane models.

The following significant flaws in DGP\footnote{Without including the DE component, the DGP model provides a straightforward mechanism for cosmic acceleration; however, it is unable to adequately fit the data. First off, the expansion history anticipated by the DGP model seems to contradict the data from SNIa, BAO, and the CMB, indicating this model is unacceptably inaccurate \cite{Davis:2007na}. Rubin et al. demonstrated in Ref. \cite{Rubin:2008wq} that the DGP model leads to a $\Delta_ \text{min}\chi^2=15$ compared to the $\Lambda$CDM model using the Union+BAO+CMB data. Second, data does not support the DGP model's history of structure formation \cite{Song:2006jk,Schmidt:2009sg}. Fang et al. shown in Ref. \cite{Fang:2008kc} that the low-$l$ anisotropies of the CMB temperature power spectrum are over-predicted by this model. DGP is excluded at 4.9$\sigma$ (with curvature) and 5.8$\sigma$ (without curvature) levels in comparison to the $\Lambda$CDM. Consequently, when compared to cosmological observations, the DGP model's condition is far from favorable \cite{Guo:2006ce,Fairbairn:2005ue,Maartens:2006yt,Xia:2009gb}. The DGP model includes a scalar ghost field localized close to the brane, in accordance with investigations of the linear theory concerning a flat multidimensional space-time and a flat brane \cite{Dubovsky:2002jm}.
Nevertheless, by embedding our visible $3D$ brane within a $4D$ brane in a flat $6D$ bulk, we may obtain a ghost-free DGP model \cite{deRham:2007xp}. Galilean gravity also offers an opportunity to get around the ghost problem \cite{Nicolis:2008in}. It is crucial to recall that including the inflationary phase in the model can help to ease some of the concerns highlighted above \cite{Zhang:2004in,Garcia-Aspeitia:2020snv}.}, and RS (II) models led to the development of the covariant extrinsic gravity (for more details, please see Ref. \cite{Jalalzadeh:2013wza}): 1) {These models have phenomenological characteristics.} 2) The majority of the innovations are unique to individual models. For instance, this has misled some into thinking that the brane theory is a $5D$ theory based on the $AdS5$ or Ricci flat bulk. 3) The geometric process for unifying the fundamental forces utilizing a group of non-compact bulk space isometrics still needs to be fully established, unlike the KK and multidimensional gravitational models. 4) The brane has been assumed to be extremely thin as a means of simplification. As long as the energy scales are substantially less than the energy scales associated with the inverse thickness of the brane or at distance scales far larger than the thickness of the brane, this approximation is presumed to be accurate. 5) The fact that a junction condition is not unique is the major challenge in implementing one. There are more types of junction conditions. Therefore different junction conditions might have various physical effects. Additionally, these junction conditions are inapplicable when several non-compact extra dimensions are present.

The Extrinsic gravity (or Covariant gravity) approach is described in Refs. \cite{Jalalzadeh:2013wza,Rostami:2015ixa} is the main focus of the current paper because it is free of the aforementioned issues. These references derive covariant Einstein's equations for a braneworld model that is locally and isometrically embedded in a bulk space with any number of dimensions. The brane is given a thickness thanks to the application of Nash's theorem to the perturbation of the submanifold. Usually, the junction condition is used to define the extrinsic curvature in brane models with one non-compact additional dimension. The junction condition is not relevant in the case of multiple extra dimensions \cite{1986GReGr695M,1994PhRvD7233M,2002IJMPA341M,2002PhLA9M,2011GReGr2685M,Jalalzadeh:2013wza,Rostami:2015ixa}, though.

The structure of this article is as follows. We review the model's essential geometric characteristics in section \ref{Geometry}. The field equations for the 3-brane, with thickness $l$, embedded in a higher-dimensional space-time, where gravity lives in the bulk and matter fields and gauge fields in the brane, are driven in section \ref{Feild}. Section \ref{Brane} presents the Bianchi type-V as a prototype model, which includes FLRW as a special case, braneworld cosmology. 
Because there is a slight difference in the intensities of the microwaves received from different directions, Hinshaw et al. \cite{2013ApJS19H} have proposed that the metric components may be of distinct functions of time. It is necessary to take into account space-time that behaves somewhat like the FLRW metric in order to compare the detailed observations. It could be better to use the spatially anisotropic but homogeneous Bianchi type-V space-time for this.
We obtain modified Friedmann and Raychaudhuri equations in section \ref{FR} for the model. Observational constraints are discussed in section \ref{Obser}, and they are used to determine the number of extra dimensions and the extrinsic tube in section \ref{Prob}. Also, we show how the number of extra dimensions solves the coincidence problem. In section \ref{Wave}, we show that the model predicts a mass for gravitational waves. In addition, we discuss Zeldovich, Weinberg, and Wesson’s numerical relations and show how our model explains these relations. Section \ref{Con} contains our conclusions and a thorough
discussion. 

For the remainder of this paper, we shall work in natural units, $\hbar=c=k_\text{B}=1$.

\section{	The model's geometrical aspects }\label{Geometry}
This section reviews the basic mathematical properties of extrinsic gravity proposed in Refs. \cite{Jalalzadeh:2013wza,Rostami:2015ixa}, where our Universe is embedded in a $D$-dimensional flat bulk space.   Consider a $4D$ space-time manifold $(\mathcal M_4,\bar g)$  with local coordinate patch, $x^\mu,~~\mu=0,1,2,3$, embedded locally and isometrically in a $D$-dimensional Bulk space $(\mathcal M_D,\mathcal G)$, with a natural Cartesian coordinates $\mathcal Y^A,~~A=0,...,D-1$.

According to the embedding requirement, a confined observer in a 3-brane who uses just the induced metric as his primary tool must agree with a Gaussian observer whose observations require the second fundamental form to compute $\mathcal M_4$ in $\mathcal M_D$. For such an agreement, the extrinsic curvature $\bar K_{\mu\nu}$ and the compatibility equations between the Riemannian and Gaussian geometries must have physical interpretations.  This allows us to construct an adapted
coordinate patch in bulk, which includes the local coordinates of submanifold $\mathcal M_4$ as $\{x^\mu, x^a\}$, where
$x^a:=\{x^4,...,x^{D-1}\}$ are extrinsic (or extra) coordinates. Then, the submanifold is defined by $x^a=0$.
The isometric local embedding is given by $D$ differential maps
\begin{equation}\label{1-1}
    \mathcal Y^A : \mathcal M_4 \rightarrow \mathcal M_D.
\end{equation}
The embedding equations are given by
\begin{equation}
\begin{split}
    \mathcal{G}(\Bar{e}_\mu,\Bar{e}_\nu)&=\mathcal{G}_{AB}Y^A_{,\mu}Y^B_{,\nu}=\Bar{g}_{\mu\nu},\\
   \mathcal{G}(\Bar{e}_\mu,\Bar{e}_a)&=\mathcal{G}_{AB}Y^A_{,\mu}N^B_a=0,\\
   \mathcal{G}(\Bar{e}_a,\Bar{e}_b)&=\mathcal{G}_{AB}N^A_{a}N^B_{b}=\delta_{ab},
   \end{split}
\end{equation}
which
\begin{equation}
\Bar{e}_\mu:=Y^A_{,\mu}\partial_A,~~~~
\Bar{e}_a:=N^A_{a}\partial_A ,
\end{equation}
 are tangent and normal vector spaces for each point of $4D$ space-time, and $\delta_{ab}$ is the Euclidean metric of the orthogonal space.
 Also, $N^A_a$ denotes the components of $\mathfrak N=D-4$ unit vector fields orthogonal to the $\mathcal M_4$
and also normal to each other in the direction of the extra coordinates. 

The perturbation of $\mathcal M_4$ in an adequately small neighborhood
of the brane along an arbitrary transverse direction, $\eta$  is given by \cite{Maia:1984nv}
\begin{multline}
    Z^A(x^\mu ,\eta)=Y^A(x^\mu)+\zeta({\mathcal L_\eta}Y)^A=\\Y^A(x^\mu)+\zeta[\eta,Y]^A,
\end{multline}
where $\mathcal L$ is the Lie derivative, and the perturbation is applied along an arbitrary direction $\eta$, which is parameterized by $\zeta$. The tangential projection of $\eta$ can always
be identified with the action of brane diffeomorphism, which will subsequently be ignored. Also, one can insert additional simplification by taking the norm of $\eta$ equal
to unity. Hence, we can consider a general deformation along the normal vectors $\eta=\bar e^a$
parameterized by extra dimensions $x^a$. Consequently perturbation along orthogonal
direction $\bar e^a$ will be
\begin{equation}
  Z^A_{,\mu}(x^\alpha ,x^a)=Y^A_{,\mu}(x^\alpha)+x^a{N}^A_{a,\mu}.
\end{equation}
Similarly, because the vectors $N^A$ depend just on the local coordinates $X^\mu$, they do not propagate along the extra dimensions. As a result, the perturbed embedding will be 
\begin{equation}
    Z^A_{,m}=N^A_m.
\end{equation}
The above equations show that the deformed embedding
will be
\begin{equation}\label{sh6}
    \begin{split}
 Z^A_{,\mu}& = Y_{,\alpha}^A(\delta^\alpha_\mu-x^a\bar K_{\mu}^{~\alpha a})+N^A_mA_{\mu m}, \\
  Z^A_{,m}&=N^A_m,
    \end{split}
\end{equation}
where $A_{\mu a}=J^{mn}_aA_{mn\mu}$, and $J^{mn}_a$ are the killing vector basis of the space generated by extra dimensions given by $J^{mn}_a=x^{[m}\delta^{n]}_a$ \cite{Maia:1983zh}.
If we take ${N^A}_m={\delta^A}_m$, the bulk space metric, parametrized by Gaussian coordinates $\{x^\mu,x^a\}$, can be represented in the matrix form shown below
\begin{equation}\label{sh7}
  \mathcal{G}_{AB}=
\begin{pmatrix}
g_{\mu\nu}+\lambda^2 A_{\mu c}A^{c}_\nu & \lambda A_{\mu j} \\
\lambda A_{\nu i} & \delta_{ij}
\end{pmatrix},
\end{equation}
where
\begin{equation}\label{sh8}
\begin{split}
    g_{\mu\nu}&=\Bar{g}_{\mu\nu}-2x^k \Bar{K}_{\mu\nu K}+ x^m x^n \Bar{K}_{\mu \alpha m}\Bar{K}^{\alpha}_{\nu n}\\
    &=\Bar{g}^{\alpha\beta}(\Bar{g}_{\mu\alpha}-x^a \Bar{K}_{\mu\alpha a})(\Bar{g}_{\nu\beta}-x^b \Bar{K}_{\nu\beta b}),
    \end{split}
\end{equation}
 is the metric of the brane with thickness $l$ \cite{Jalalzadeh:2013wza,Maia:1983zh}. Note also that we rescaled the gauge potential $A_{\mu a}$ to $\lambda A_{\mu a}$, where $\lambda$ is a scaling factor we shall choose later so that rescaled gauge potential is a conventionally normalized gauge field.

\begin{figure}[h!]
  \centering
  \includegraphics[width=8cm]{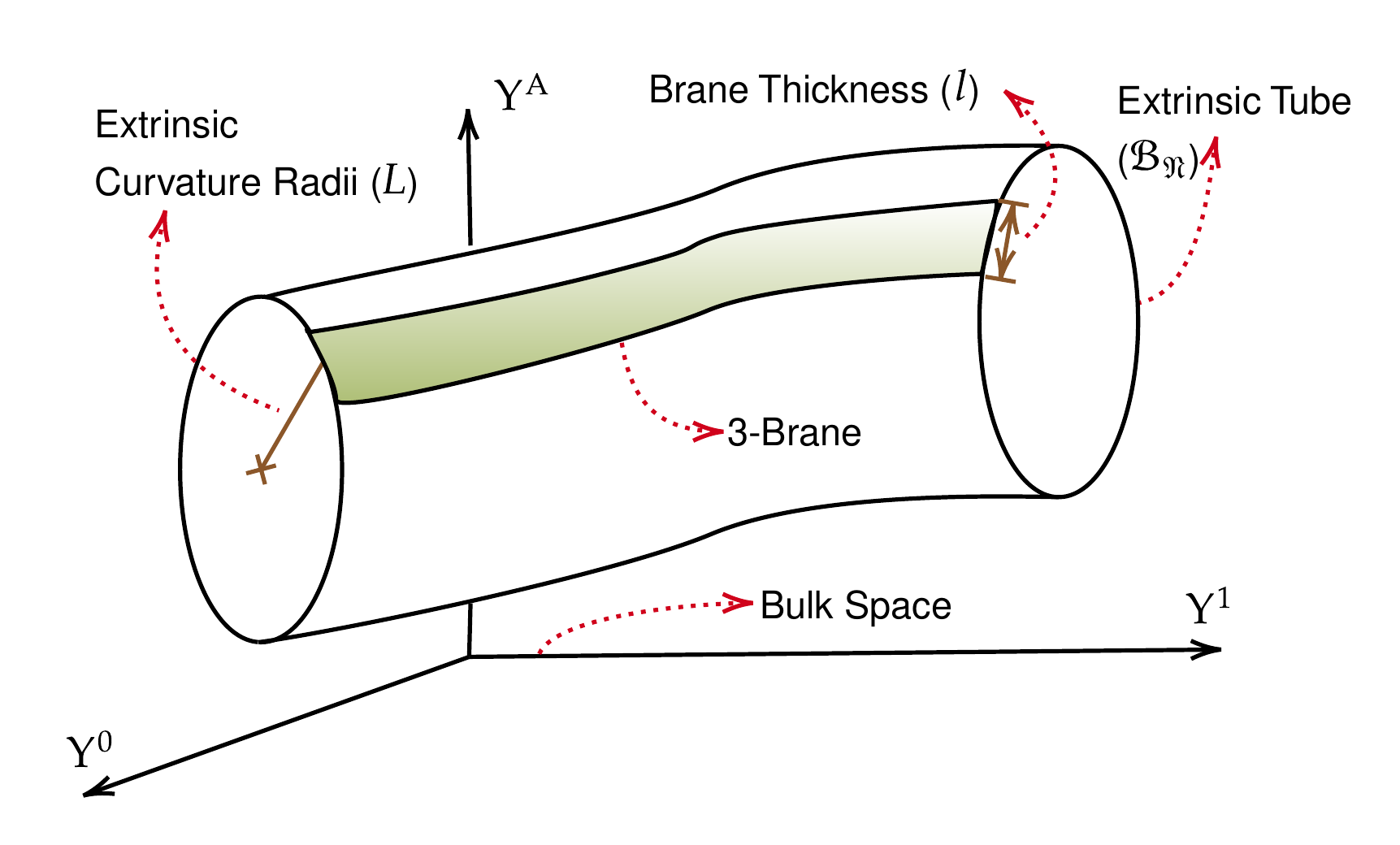}
  \caption{\small  Brane of thickness $l$ is embedded in $D$-dimensional bulk, which is bounded by an Extrinsic tube generated by the normal curvature radii $L$. Matter particles are confined to a line of length $l$ in each point of the brane in this figure for illustration.}\label{fig}
\end{figure}

 Regarding the Gaussian form of the metric of the bulk space, given by (\ref{sh7}), $\det\mathcal{G}_{AB}=\det g_{\mu\nu}\det\delta_{ij}\neq0$. Thus, from (\ref{sh8}), $x^a$ must not satisfy $\det(\bar g_{\alpha\beta}-x^a\bar K_{a\alpha\beta})=0$. The last equation defines the curvature radii, $L_a$,  of $\mathcal M_4$ associated with the  the principal directions $\delta x^\mu$ and each normal vector $\bar e_a$ \cite{Eisenhart}
 \begin{equation}\label{1-18}
    (\Bar{g}_{\mu\alpha}-L^a\Bar{K}_{\mu\alpha a})\delta x^\mu =0.
\end{equation}
In the vicinity of any of its points, the brane's extrinsic curvature is identified using normal curvature, defined by
\begin{equation}
   \frac{1}{R_a}=\frac{\bar K_{\alpha\beta a}\delta x^\alpha \delta x^\beta}{\bar g_{\alpha\beta a}\delta x^\alpha \delta x^\beta}.
\end{equation}
It follows that the metric of the deformed submanifold (\ref{sh8}) becomes singular at the solution of equation (\ref{1-18}).
Hence, at each point of the submanifold, the normal curvature radii generate a closed space ${\mathcal B}_{\mathfrak N}$ (Extrinsic tube). The physical space on the ambient space is bounded to the Extrinsic tube as shown in Fig. \ref{fig}. All extra dimensions
are assumed to be spacelike, then ${\mathcal B}_{n}$ may be taken locally to be the $n$-sphere
$S_{n}=SO(\mathfrak N)/SO(\mathfrak N-1),$
with radius  $L:=min\{L^a_{(\mu)}\}$, at each point of space-time \cite{Maia:1984nv,Bueno:2022log}.
According to the confinement hypothesis, it is supposed that the standard model particles are localized to a $4$-dimensional submanifold, while gravity can freely propagate in the ambient space. However, due to equation (\ref{1-18}), the normal curvature radii determine a characteristic radius, for the propagation of the gravitons.
Note that Eq. (\ref{1-18}) admits a nontrivial solution for curvature directions $\delta x^\mu$ when
 \begin{equation}\label{1-18a}
    \det(\Bar{g}_{\mu\alpha}-L^a\Bar{K}_{\mu\alpha a}) =0.
\end{equation}

 According to the (\ref{sh6}), the reference frames of the perturbed geometry are defined as
\begin{equation}
    h^A_{\mu}=Y^A_{,\alpha}(\delta^\alpha _ \mu - x^a \Bar{K}^\alpha _{\mu a}),~~~~
h^A_{a}=N^A_{a},
\end{equation}
and the perturbed brane's tangent and normal vectors, respectively, by
\begin{equation}
\begin{split}
    e_{\alpha}&=\partial_\alpha -\lambda A^m_\alpha \partial_m = h^A_\alpha \partial_A, \\                e^\alpha &= dx^\alpha, \\
e_m&= \partial_m = N^A_m \partial_A,\\
e^m&= dx^m +\lambda A^m_\alpha dx^\alpha.
\end{split}
\end{equation}
As a result, the perturbed brane's embedding equations will be
\begin{equation}
\begin{split}
    \mathcal{G}(e_\alpha,e_\beta)&=\mathcal{G}_{AB}h^A_{\alpha}h^B_{,\beta}=g_{\alpha\beta} ,\\
  \mathcal{G}(e_\alpha,e_m)&=\mathcal{G}_{AB}h^A_\alpha N^B_m=0,\\
  \mathcal{G}(e_m,e_n)&=\mathcal{G}_{AB}N^A_m N^B_n=\delta_{mn}.
  \end{split}
 \end{equation}

Generally, by substitution of ${ Z}^A_{,\mu}$ derived from (\ref{sh6})
into the  extrinsic curvature of deformed submanifold, defined by $K_{\mu\nu
a}(x^\alpha,x^b):=-{\mathcal G}_{AB}{ N}^A_{;\mu}{ Z}^B_{,\nu}$, we
obtain the extrinsic curvature of a deformed submanifold in terms of the extrinsic curvature of the original non-perturbed submanifold as
\begin{multline}\label{sha1}
K_{\mu\nu a}(x^\alpha,x^b)=\\
K_{\mu\nu a}(x^\alpha)-\frac{1}{2}x^m\left(K_{m\gamma(\mu }K_{\nu)\,\,\,\,a}^{\,\,\,\,\gamma}-\lambda F_{\mu\nu an}\right).
\end{multline}
Comparing definition of $\gamma_{\mu\nu}$ in (\ref{sh8}) with (\ref{sha1})
we obtain
\begin{eqnarray}\label{sha2}
K_{\mu\nu a}(x^\alpha,x^b)=-\frac{1}{2}\partial_a\gamma_{\mu\nu}+\frac{\lambda}{2}x^bF_{\mu\nu
ab},
\end{eqnarray}
where $F_{ab\mu\nu}$ is the curvature associated with extrinsic twist vector field $A_{\mu ab}$, defined as  \cite{twist}
\begin{eqnarray}
\label{1-10}
F_{ab\mu\nu}=A_{\mu ab,\nu}-A_{\nu ab,\mu}-A_{\nu a}^{\,\,\,\,\,\,\,c}A_{\mu
cb}+A_{\mu a}^{\,\,\,\,\,\,\,c}A_{\nu cb},
\end{eqnarray}
where $A_{\mu ab}$ plays the role of the Yang--Mills potential \cite{Yang}. Notice that $A_{\mu ab}$ transform as the component of a gauge vector field under the group of isometries of the bulk if the bulk space has certain
Killing vector fields \cite{Shahram}, and it only exists in the ambient spaces with dimensions equal to or
greater than six $(\mathfrak N\geqslant 2)$.
The extrinsic curvature gives a measure
of the deviation from the submanifold and its tangent plane at any point.
Its symmetric part shows that the second fundamental form
propagates in bulk. The antisymmetric part is proportional to a Yang--Mills
gauge field, which really can be thought of as a kind of curvature.

\section{Effective action on the brane and field equation}\label{Feild}

As it is demonstrated in detail in Ref. \cite{Jalalzadeh:2013wza}, the above new embedding helps us to rewrite the Ricci scalar of the bulk space, $\mathcal R$ in terms of the Ricci scalar of the original brane, $\bar R$, the extrinsic curvature, $\bar K_{a\mu\nu}$, and the Yang--Mills potential $F_{\mu\nu ab}$
\begin{multline}\label{31}
\mathcal{R}=\bar{R}+\bar{K}_{\alpha\beta a}\bar{K}^{\alpha\beta a}-\bar{K}_a\bar{K}^a+\\x^m\Big\{\bar{g}^{\gamma\beta}\nabla^{(tot)}_\beta
H^\sigma_{\sigma\gamma m}
-\bar{g}^{\gamma\beta}\nabla^{(tot)}_\sigma H^\sigma_{\beta\gamma
m} +\\2\bar{K}^{\alpha\beta}_{\,\,\,\,m}\bar{R}_{\alpha\beta}\Big\}-\frac{\lambda^2}{4}x^mx^nF^{\alpha\beta}_{\,\,\,\,\,\,am}F_{\alpha\beta\,\,\,\,\,\,n}^{\,\,\,\,\,\,\,\,a},
\end{multline}
where $\bar K_a=\bar g^{\mu\nu}\bar K_{\mu\nu a}$, $\nabla_\mu^{(tot)}$ is the total covariant derivative,  defined by
\begin{eqnarray}\label{1-27}
\nabla_\mu^{(tot)}{\bar K}_{\alpha\beta m}={\bar K}_{\alpha\beta m;\mu}-\lambda A_{\mu
mn}{\bar K}_{\alpha\beta}^{\,\,\,\,\,\,n},
\end{eqnarray}
and
\begin{equation}
    H^\rho_{\alpha\beta
m}=\bar{g}^{\rho\mu}\left(\nabla_\alpha^{(tot)}\bar{K}_{\mu\beta
m}+\nabla_\beta^{(tot)}\bar{K}_{\mu\alpha m}-\nabla_\mu^{(tot)}\bar{K}_{\alpha\beta
m}\right).
\end{equation}
Now, using expression (\ref{sh8}) and Eq. (\ref{31}) in Einstein--Hilbert action functional of the bulk, and remembering the confinement
hypothesis (matter and gauge fields are confined to the brane with thickness $l$
while gravity could propagate in the bulk space up to the curvature radii $L$ with
the local geometry of $\mathbb S^{\mathfrak N}$) we obtain
\begin{multline}\label{1-30004}
-\frac{1}{2\kappa^2_D}\int{\sqrt{|\mathcal{G}|}{\cal R}d^Dx}=\\
-\frac{M_D^{\mathfrak N+2}V_{\mathfrak N}}{16\pi}\int L^{\mathfrak N}
\sqrt{|\bar{g}|}\Big(\bar{R}-\bar{K}_{\alpha\beta m}{\bar K}^{\alpha\beta m}
+ \bar{K_a}\bar K^a\Big)d^4x\\
-\frac{dM_D^{{\mathfrak N}+2}V_{{\mathfrak N}-1}\lambda^2}{48\pi}\int L^{{\mathfrak N}-1}l^3\sqrt{|\bar g|}\Tr (F_{\mu\nu}F^{\mu\nu})d^4x,
\end{multline}
where $M_D$ is the fundamental energy scale in the bulk space, $V_{\mathfrak N}=\pi^{{\mathfrak N}/2}/\Gamma({\mathfrak N}/2+1)$, and $d=2,4,10$, for the Maxwell, the Weak, and the strong interactions, respectively. Therefore, the induced action functional is equivalent to the $4D$ gravitational action containing extrinsic terms and Maxwell--Yang--Mills action.
To obtain the $4D$ gravitational constant, $G_N$ in terms of the fundamental scale, $M_D$, is given by
\begin{eqnarray}\label{sh11}
\frac{1}{16\pi G_N}=\frac{\pi^{\frac{{\mathfrak N}}{2}-1}}{16\Gamma(\frac{{\mathfrak N}}{2}+1)}L^{\mathfrak N}M_{D}^{{\mathfrak N}+2}.
\end{eqnarray}

Also, to obtain standard normalization for the gauge fields, the Lagrangian density of the Maxwell--Yang--Mills fields
\begin{equation}
    \mathcal L_\text{MYM}=-\frac{1}{4}\Tr(F_{\mu\nu}F^{\mu\nu}),
\end{equation}
we choose
\begin{equation}
    \label{Lambda}
    \lambda^2=\frac{12\pi^\frac{3}{2}\Gamma(\frac{{\mathfrak N}+1}{2})}{d\Gamma(\frac{{\mathfrak N}+2}{2})}\frac{LL_\text{Pl}^2}{l^3}.
\end{equation}

{Regarding the confinement hypotheses, particles, and fields are confined to the brane with thickness $l$. Concerning quantum mechanics, this implies that the wave function of a particle along the extra dimensions is confined up to a distance $l$. In other words, because the uncertainty principle prohibits the wave function of a test particle from being precisely localized on the brane at the quantum level, the gauge fields (with geometric origin) and extrinsic curvature effects become pronounced in the Klein--Gordon (KG) equation induced on the brane. This means that the mass spectrum of the particle should be related to the thickness $l$. Therefore, one should expect to relate the thickness (a geometric quantity) to the mass of particles. In addition, regarding the geometric origin of gauge fields in our model, the induced KG should give us the parameter $\lambda$ in terms of gauge couplings. }                         

{Let's start with the massless KG equation in the bulk space 
\begin{equation}
    \label{Gordon1}
    \mathcal G^{AB}\nabla_A\nabla_b\psi(x^\mu,x^a)=0,
\end{equation}
 where $\psi$ is the wave function of the particle with the normalization condition 
 \begin{equation}
     \int \sqrt{|\mathcal G|}|\psi|^2d^Dx=1.
 \end{equation}
 We rescale the wave function so that it is normalized in $L^2(\mathcal M_4)$ since our aim is to have efficient dynamics on the brane. The following scalings \cite{Jalalzadeh:2004uv} are used to accomplish this goal:
 \begin{equation}\begin{split}
  &   \Phi(x^\mu,x^a)=\left(\frac{|\bar g|}{|\mathcal G|}\right)^\frac{1}{4}\psi(x^\mu,x^a),\\
&     \mathcal G^{AB}\nabla_A\nabla_b\rightarrow\left(\frac{|\bar g|}{|\mathcal G|}\right)^\frac{1}{4}\mathcal G^{AB}\nabla_A\nabla_b\left(\frac{|\mathcal G|}{|\bar g|}\right)^\frac{1}{4}.
     \end{split}
 \end{equation}
   Splitting the tangent and normal degrees of freedom, $\Phi(x^\mu,x^a)=\sum\Phi_1(x^\mu)\Phi_2(x^a)$, 
gives us \cite{Jalalzadeh:2004uv} the following equations}
\begin{multline}\label{Gordon2}
 \Big\{   -\frac{1}{\sqrt{|\bar g|}}(\partial_\mu-i\lambda A_\mu)\bar g^{\mu\nu}\sqrt{|\bar g|}(\partial_\nu-i\lambda A_\nu) +\\
 \frac{1}{4}(\bar K_a\bar K^a-2\bar K_{\alpha\beta a}\bar K^{\alpha\beta a})+m^2\Big\}\Phi_1(x^\sigma)=0,
\end{multline}
and
\begin{equation}\label{Gordon3}
   \big\{\delta^{ab}\partial_a\partial_b +m^2\big\}\Phi_2(x^m)=0,
\end{equation}
where $m$ is the induced mass. The last equation gives a mass spectrum of the $4D$ fields in (\ref{Gordon2}). Regarding that, the matter particles are confined within a brane with width $l$ (in a good approximation, one can assume the particle is confined inside a $\mathfrak N$-dimensional box with a side equal $l$, see Fig. \ref{fig}.); the first excited state of the mass spectrum in Eq. (\ref{Gordon3})  gives us the relation between the lightest particle with the width of the brane.
\begin{equation}
    \label{Gordon4}
    m=\frac{\sqrt{{\mathfrak N}}\pi}{l}.
\end{equation}
The fields $\Phi_{2}$ are thus the mass eigenstates in the brane. Eq. (\ref{Gordon2}) shows that the field $\Phi_{1}$ has charge $g=\lambda$. Therefore,
\begin{equation}
    \label{Gordon5}
  \frac{g^2}{4\pi}=\frac{3\sqrt{\pi}\Gamma(\frac{{\mathfrak N}+1}{2})}{d\Gamma(\frac{{\mathfrak N}+2}{2})}\frac{LL_\text{Pl}^2}{l^3}.
\end{equation}

Eq. (\ref{Gordon5}) immediately gives us the following fundamental relation between the normal
curvature radii, the thickness of the brane, the number of extra dimensions, and $4D$ Planck's length ($M_\text{Pl}=1/\sqrt{G_N}=1/L_\text{Pl}$)
\begin{eqnarray}\label{New11}
L=\frac{d\Gamma(\frac{{\mathfrak N}+2}{2})}{3\sqrt{\pi}\Gamma(\frac{{\mathfrak N}+1}{2})}\frac{g^2}{4\pi}\left(\frac{l}{L_\text{Pl}} \right)^3L_\text{Pl}.
\end{eqnarray}

Note that if $ L\sim l$, the above equation reduces to the equivalent relation
in KK gravity \cite{Love}. Also, from equation (\ref{New11}),
it is easy to see that the following relation is held between gravitational
``constant'' and gauge couplings
\begin{eqnarray}\label{r1}
\frac{G_N}{G_0}=\left(\frac{g^2}{g^2_{0}}\right)^{-\frac{{\mathfrak N}}{{\mathfrak N}+1}},
\end{eqnarray}
where $G_0$ and $g_{0}$ are the gravitational constant and the gauge coupling constants
at the present epoch, respectively.

Now, the induced Einstein--Yang--Mills field equations for the thick braneworld model will be \cite{Jalalzadeh:2013wza}
\begin{equation}
\label{1-38a}
{G}_{\alpha\beta}= -Q_{\alpha\beta}+8\pi G_N\left(T_{\alpha\beta}+T_{\alpha\beta}^\text{(YM)}\right),
\end{equation}
\begin{equation}
\label{1-38b}
\nabla^{(tot)}_\beta{K}_a-\nabla^{(tot)}_\alpha{K}_{\beta a}^{\alpha}=8\pi
G_NT_{a\beta},
\end{equation}
\begin{multline}
\label{1-38c}
\frac{G_N}{g_i^2}\left(F^{\alpha\beta}_{\,\,\,\,\,\,\,am}F_{\alpha\beta
b}^{\,\,\,\,\,\,\,\,\,\,m}+\frac{1}{2}\eta_{ab}F_{\alpha\beta}^{\hspace{.3cm}lm}F^{\alpha\beta}_{\hspace{.3cm}lm}\right)-
\\
-\frac{1}{2}\delta_{ab}\left({R}+{K}_{\mu\nu m}{K}_{\mu\nu}^{\hspace{.3cm}m}-{K}_a{K}^a\right)=8\pi
G_NT_{ab},
\end{multline}
where ${G}_{\alpha\beta}$ is the $4D$ Einstein tensor and  $G_{N}$ is the induced gravitational constant.
$T_{\alpha\beta}$, $T_{a\beta}$, and $T_{ab}$ are the components of the energy-momentum tensor defined such that to be compatible with the confinement hypothesis.
$T_{\alpha\beta}^\text{(YM)}$ denotes the Yang--Mills energy-momentum tensor and $Q_{\alpha\beta}$ is a conserved quantity
expressed in terms of extrinsic curvature and its  trace, $K_a$, as
\begin{multline}\label{1-0}
Q_{\alpha\beta}={K}_{\alpha}^{\,\,\,\,\eta a}{K}_{\beta\eta a}-{K}^a{K}_{\alpha\beta a}-\\ \frac{1}{2}{g}_{\alpha\beta}({K}^{\mu\nu
a}{K}_{\mu\nu a}-{K}_a{K}^a).
\end{multline}
Note that the tensor
$Q^{\alpha\beta}$ is conserved in the sense that $\nabla_\beta Q^{\alpha\beta}=0$, as it can be directly
verified \cite{Maia:2004fq}.

The resulting field equations provide both the equations of general relativity and of Yang--Mills. Thus, it introduces a unifying picture like the KK theory \cite{Kaluza:1921tu}.
  
\section{ Field equations on the Bianchi type-V braneworld}\label{Brane}  

{The large-scale structure's isotropic and homogenous properties might represent an asymptotic circumstance arising from the Universe's anisotropic origin, which is generated by the matter component during decoupling. In order to understand how the anisotropy parameters change as the Universe progresses into the present epoch, it is crucial to construct an isotropization criterion.}

Let us examine how a homogeneous and anisotropic universe is affected by extrinsic curvature terms. We assume that the Bianchi type-V  space-time is embedded locally and isometrically in a $D = {\mathfrak N} + 4$ dimensional Minkowskian
space-time, in which all extra dimensions are spacelike. The contribution of weak and strong interactions in the Universe's history (the age, the amount, the density parameters,  etc.) are ignorable. Also, the density parameter of the radiation at the present day, $\Omega^\text{(radiation)}_0$, is of the
order of $10^{-5}-10^{-4}$, radiation becomes essential only for high
redshifts ($z\gtrsim1000$). Thus, we also assume that the twisting vector fields, $A_{\mu a b}$, vanish for simplicity.

The space-time metric of the spatially homogeneous and anisotropic Bianchi types-I and V is given by
    \begin{multline}\label{3-1}
      ds^2=-dt^2+a_1(t)^2dx^2+\\a_2(t)^2e^{2\beta x}dy^2+a_3(t)^2e^{2\beta x}dz^2,
  \end{multline}
  
The Bianchi type-I and V geometries are represented by the instances $\beta=0$ and $\beta=1$, respectively. The scale factors in different spatial directions are denoted by $\{a_1,a_2,a_3\}$. Let us define some  physical and geometrical parameters that will be utilized in establishing the law and solving field equations for the metric (\ref{3-1}).
One can define the average scale factor, $a(t)$, as
\begin{equation}\label{3-2}
    a(t)=\Big(a_1(t)a_2(t)a_3(t)\Big)^\frac{1}{3}.
\end{equation}
{This relation shows that  the total expansion factor $a(t)$ has the contribution from
each directional expansion factor.}
A volume scale factor, $ V$, corresponding to the average scale factor is defined by
\begin{equation}\label{3-3}
    V=a(t)^3=a_1(t)a_2(t)a_3(t). 
\end{equation}
Also, one can define the generalized mean Hubble’s parameter, $H$, 
\begin{equation}\label{3-4}
    H=\frac{1}{3}(H_1+H_2+H_3),
\end{equation}
where $H_1=\dot a_1/a_1$, $H_2=\dot a_2/a_2$ and $H_3=\dot a_3/a_3$ are the directional Hubble’s parameters in the directions of $x$, $y$ and
$z$ respectively. These definitions lead us to the following relation
\begin{equation}
    \label{3-6}
    H=\frac{\dot a}{a}=\frac{1}{3}\frac{\dot V}{V}=\frac{1}{3}(H_1+H_2+H_3).
\end{equation}
In cosmology, the physical quantities of observational relevance are the expansion scalar, $\Theta$, the shear scalar $\sigma^2$, and the average anisotropy parameter, $A_p$. All of them are defined as follows, respectively
\begin{equation}
    \label{3-7}
    \begin{split}
        \Theta&=H_1+H_2+H_3=3H,\\
        \sigma^2&=\frac{1}{2}\left(H_1^2+H_2^2+H_3^2\right)-\frac{\Theta^2}{6},\\
        A_p&=\frac{1}{3}\sum_{\alpha=1}^3\left(\frac{\delta H_\alpha}{H}\right)^2,
    \end{split}
\end{equation}
where $\delta H_\alpha=H_\alpha-H$. {The average anisotropy is a measure of divergence from isotropic expansion. When $A p=0$, the model's isotropic behavior may be obtained. The shear tensor, $\sigma_{ij}$,  reflects any propensity for the initially spherical area to deform into an ellipsoidal shape. The region's rate of distortion is therefore represented by the shear scalar, $\sigma^2$.}

The energy-momentum tensor in field equations (\ref{1-38a}), (\ref{1-38b}) and (\ref{1-38c}) is diagonal due to the symmetries of the embedded $4D$ Bianchi braneworld. Also, we investigate the model in the absence of anisotropic sources. As a result, we use a perfect fluid form for the energy-momentum tensor in comoving coordinates, in which $p_x=p_y=p_z=p$. Thus, we have
  \begin{equation}\label{3-8}
  \begin{split}
      T_{\mu \nu}&=(\rho +p)u_\mu u_\nu +pg_{\mu \nu},~~~~u_\mu=-\delta_\mu^0,\\
       T_{\mu a}&=0,\\ 
       T_{ab}&=p_{ext} \eta _{ab},
\end{split}      
  \end{equation}
where $u^\mu$ denotes the 4-velocity of the confined perfect fluid, $\rho$ is the confined matter density,  $p$ is the corresponding pressure, and $p_{ext.}$ is the pressure along the extra dimensions, inside of thick brane. Note that by determining the trace of Eq. (\ref{1-38a}), which gives the Ricci scalar of the brane, and
plugging it in Eq. (\ref{1-38c}), we obtain $p_{ext.}=(3p-\rho)/2$. Therefore, the pressure along the extra dimensions is not an independent thermodynamical variable, and it depends on the type of confined perfect fluid.

  To determine the induced Einstein's field equations, first, we need to obtain the components of $Q_{\alpha\beta}$ defined by Eq. (\ref{1-10}). 
Regarding the York’s relation (\ref{sha2}),
 for a diagonal metric $K_{\mu \nu a}$ is also diagonal. Thus, $K_{~\nu a}^\mu=\text{diag}(K_{~0 a}^0, K_{~1 a}^1,K_{~2 a}^2,K_{~3 a}^3)$. Regarding the above assumptions, the field equations (\ref{1-38b}) turn out
 \begin{equation}
     \label{3-9}
 \nabla_\beta{K}_a-\nabla_\alpha{K}_{~\beta a}^{\alpha}=0,  
 \end{equation}
where $\nabla_\alpha$ is the induced covariant derivative of the brane, and $K_a=g^{\alpha\beta}K_{\alpha\beta a}=K^0_{~0a}+K^1_{~1a}+K^2_{~2a}+K^3_{~3a}$. For values $\beta=1,2,3$  in the field equation (\ref{3-9}) we find the following equations respectively
\begin{equation}\label{3-10}
    \begin{split}
 &\partial_1(K_a-K^1_{~1a})+\beta(2 K^1_{~1a}- K^2_{~2a}- K^3_{~3a})=0,\\
& \partial_2(K_a-K^2_{~2a})=0,\\
& \partial_3(K_a-K^3_{~3a})=0.
    \end{split}
\end{equation}
  Because of the homogeneity of space, $Q_{\mu\nu}$ in the induced Einstein's field equation can only be a function of cosmic time, $t$. This necessitates that the extrinsic curvature components only be a function of the time coordinate.
  Hence, the last term in the first equation of (\ref{3-10}) must vanish, $2 K^1_{~1a}- K^2_{~2a}- K^3_{~3a}=0$. And finally, considering $\beta=0$ in Eq. (\ref{3-9}) and using the last relation between the spatial components of extrinsic curvature, we find
  \begin{equation}
      \label{3-11}
      K^0_{~0a}=\frac{1}{H}\dot K^1_{~1a}+K^1_{~1a},
  \end{equation}
  where a dot denotes a time derivative. By assuming that $K^2_{~2a}=K^3_{~3}=f_a(t)$, we summarize 
  \begin{equation}\label{3-12}
   \begin{split}
     K_{\alpha\beta m}&=f_m(t)g_{\alpha\beta}, ~~~~\alpha,\beta=1,2,3,  \\
     K_{00m}&=-\frac{1}{H}\dot f_{m}(t)-f_{m}(t)=-\frac{1}{aH}\frac{d}{dt}(af_m).
   \end{split}   
  \end{equation}

  At once, one can easily find the curvature radii of the model using the eigenvalue equation (\ref{1-18a})
\begin{equation}\label{3-13}
    \begin{split}
L^m_\alpha&=\frac{1}{f_m},~~~\alpha=1,2,3,\\
L_0^m&=\frac{aH}{\frac{d(a f_m)}{dt}}.
    \end{split}
\end{equation}
Regarding $\mathcal B_{\mathfrak N}$ locally is a ${\mathfrak N}$-sphere, $\mathbb S^{\mathfrak N} = SO({\mathfrak N})/SO({\mathfrak N}-1)$, one can assume that all function, $f_a,~a=1,2,...,{\mathfrak N}$, are equal, e.g., $f_a(t)=f(t)$. The normal curvature radius is given by
\begin{equation}
    \label{3-14}
    L(t)=\frac{1}{f(t)}.
\end{equation}
 According to the Eq. (\ref{sh11}), the $4D$ gravitational constant $G_N$ is time-dependent
\begin{equation}
    \label{3-15}
      G_N(t)=G_0\left(\frac{L_0}{L} \right)^{\mathfrak N}=G_0\left(\frac{f(t)}{f_0} \right)^{\mathfrak N},
\end{equation}
 where $L_0$, $f_0$ and $G_0$ are the values of $L(t)$, $f(t)$, and $G_N$  at the present epoch, $t=t_0$. The relative change in the gravitational constant thus is
\begin{equation}
    \label{3-16}
    \frac{\dot G_N}{G_N}={\mathfrak N}\frac{\dot f(t)}{f(t)}.
\end{equation}
Several constraints on the relative variation of the gravitational constant have been established using observations of Lunar laser ranging \cite{Williams:2003wu,Merkowitz:2010kka}, the Hubble diagram of distant type Ia supernovae \cite{Gaztanaga:2001fh}, the helioseismological
data of sun evolution \cite{1998ApJ871G}, or data from the binary pulsar $PSR~ 1913+ 16$ \cite{Damour:1988zz}. These findings indicate that the relative change in $G_N$ is proportional to the Hubble parameter
\begin{equation}
    \label{3-17}
    \frac{\dot G_N}{G_N} =\gamma H,~~~~|\gamma|<1,
\end{equation}
where $\gamma$ is a dimensionless constant. These constraints correspond to less than a 1\% variation of $G_N$ during the Universe's 13.7 billion-year history. The constraint (\ref{3-17}) and Eq. (\ref{3-16}) lead us to assume
\begin{equation}\label{3-18}
    \frac{h}{H}=\frac{\gamma}{{\mathfrak N}}, ~~~~|\gamma|<1,
\end{equation}
or equivalently
\begin{equation}
    \label{3-19}
    f(t)=f_0\left(\frac{a(t)}{a_0}\right)^\frac{\gamma}{{\mathfrak N}},
\end{equation}
where $a_0$ is the average scale factor at the present epoch.

Inserting the components of the extrinsic curvature obtained in (\ref{3-12}) into $Q_{\mu\nu}$ defined by Eq. (\ref{1-0}), we obtain its components 
\begin{equation}\label{3-20}
    \begin{split}
        Q_{\alpha\beta}&={\mathfrak N}f(t)^2\left(3+2\frac{h}{H}\right)g_{\alpha\beta},~~~\alpha,\beta=1,2,3,\\
        Q_{00}&=-3{\mathfrak N}f(t)^2.
    \end{split}
\end{equation}

The induced tensor $Q_{\mu\nu}$, in the sense of Wesson's Induced Matter Theory \cite{doi:10.1142/10871,doi:10.1142/6029}, should be associated with the ordinary matter as having a geometrical origin \cite{Doroud:2009zza,Jalalzadeh:2008xu,Moyassari:2007sv,Jalalzadeh:2006mr,Jalalzadeh:2006nh}. On the other hand, in Shiromizu--Maeda--Sasaki braneworld model, regarding Israel's junction
condition (and imposing the $Z_2$-symmetry) $Q_{\mu\nu}$ is proportional to the quadratic of the vacuum energy and the
energy-momentum tensor of confined matter fields \cite{Shiromizu:1999wj}.

The fact that $Q_{\mu\nu}$ is independently conserved in covariant extrinsic brane gravity, on the other hand, suggests an analogy with the energy-momentum tensor of an uncoupled non-conventional DE source \cite{Jalalzadeh:2013wza,Rostami:2015ixa,Maia:2004fq,Heydari-Fard:2006klr}.
The corresponding geometric energy-momentum of the DE is recognized to $Q_{\mu\nu}$ as
\begin{multline}
    \label{3-21}
    Q_{\mu\nu}=\\-8\pi G_N\Big\{\left(\rho^\text{(GDE)}+p^\text{(GDE)}\right)u_\mu u_\nu+p^\text{(GDE)}g_{\mu\nu}\Big\},
\end{multline}
where $u_\mu=-\delta^0_\mu$ is the 4-velocity of the ordinary perfect fluid, defined by Eq. (\ref{3-8}), and $p^\text{(GDE)}$ is the ``geometric dark energy'' (GDE) pressure associated with the extrinsic
curvature (the suffix ``GDE'' stands for extrinsic curvature, $K_{\mu\nu a}$),
and $\rho_K$ denotes the GDE density.

{In other words, in the above equation, we sought to give the extrinsic term, $Q_{\mu\nu}$, on the right side of the induced Einstein field equations (\ref{1-38a}) a form of a matter distribution (in particular, a perfect fluid). Our justification for doing this is as follows: In GR, space-time geometry is enclosed on the left-hand side of the Einstein field equation. In contrast, the distribution of matter fields in space-time appears on the right-hand side. According to this conventional viewpoint, it is common to keep this traditional viewpoint in modified gravitational field equations and to assign the form of matter distribution to the new terms that appear on the right-hand side of the field equations. Adding a CC to the gravitational field equations is the most basic example. If we write $\Lambda$ on the right-hand side of the field equations, we interpret it as a perfect fluid with negative pressure. By rewriting the extrinsic curvature term as a perfect fluid in the above equation, we have two  non-interacting perfect fluids on the matter side of Einstein's field equations. The first fluid explains the distribution of cold matter, and the second fluid has the same 4-velocity of cold matter \footnote{According to the requirements of spatial homogeneity, which imply the presence of a single comoving frame associated with observers who experience this symmetry in the surrounding Universe, both perfect fluids have the same 4-velocity \cite{Faraoni:2021opj}.}, and its origin is the bending of space-time along extra dimensions.}

Comparing the above definition of the GDE with the previous components of $Q_{\mu\nu}$ obtained in (\ref{3-20}), we find
\begin{equation}\label{3-22}
    \begin{split}
        \rho^\text{(GDE)}&=\frac{3{\mathfrak N}f^2}{8\pi G_N}=\frac{3{\mathfrak N}f_0^2}{8\pi G_0}\left(\frac{a}{a_0}\right)^{\left(\frac{2}{{\mathfrak N}}-1\right)\gamma},\\
        p^\text{(GDE)}&=-\frac{{\mathfrak N}f^2}{8\pi G_N}\left(3+2\frac{h}{H}\right)\\&=-\frac{3{\mathfrak N}f_0^2}{8\pi G_0}\left(1+\frac{2\gamma}{3{\mathfrak N}}\right)\left(\frac{a}{a_0}\right)^{\left(\frac{2}{{\mathfrak N}}-1\right)\gamma},
    \end{split}
\end{equation}
where in obtaining the second equalities, we used (\ref{3-15}) and (\ref{3-19}). One can associate an EoS with this GDE. The above forms of $\rho^\text{(GDE)}$ and $p^\text{(GDE)}$ show that
\begin{equation}
    \label{3-23}
    \omega_\text{(GDE)}=\frac{p^\text{(GDE)}}{\rho^\text{(GDE)}}=-\left(1+\frac{2\gamma}{3{\mathfrak N}}\right).
\end{equation}
This shows that the EoS of the GDE is a function of the number of extra dimensions, ${\mathfrak N}$. Regarding $|\gamma|<1$, the above equation shows that for an adequate number of extra dimensions, for example, ${\mathfrak N}=20$, $|2\gamma/(3{\mathfrak N})|\ll 1$ and as a result, the EoS of the GDE will be very near to the ``cosmological constant,'' $\omega^\text{(GDE)}\simeq -1$.

\section{Modified Friedmann and Raychaudhuri equations}\label{FR}

Let us consider the confined matter field to the brane as a perfect fluid given in comoving coordinates by (\ref{3-8}). Then, the field equations (\ref{1-38a}) 
simplify to 
\begin{subequations}
\begin{align}
 &\sum_{i<j}H_iH_j-\frac{3\beta^2}{a^2_1}=
 8\pi G_N(\rho+\rho^\text{(GDE)}),\label{4-1a}\\
& \frac{\Ddot a_2 }{a_2}+\frac{\Ddot a_3 }{a_3}+H_2H_3-\frac{\beta^2}{a^2_1}=-8\pi G_N(p+p^\text{(GDE)}),\label{4-1b}\\
 &     \frac{\Ddot a_1}{a_1}+\frac{\Ddot a_3 }{a_3}+H_1H_3-\frac{\beta^2}{a^2_1}=-8\pi G_N(p+p^\text{(GDE)}),\label{4-1c}\\
&     \frac{\Ddot a_1}{a_1}+\frac{\Ddot a_2 }{a_2}+H_1H_2-\frac{\beta^2}{a^2_1}=-8\pi G_N(p+p^\text{(GDE)}),\label{4-1d}\\
 &    2H_1-H_2-H_3=0\label{4-1e},
 \end{align}
\end{subequations}
where $H_1=\dot a_1/a_1$, $H_2=\dot a_2/a_2$ and $H_3=\dot a_3/a_3$ are the directional Hubble’s parameters defined in (\ref{3-4}).

Solving Eqs. (\ref{4-1b}), (\ref{4-1c}) and (\ref{4-1c}) yields the relationship between $a_1$, $a_2$ and $a_3$ as follows
\begin{subequations}\label{4-2}
    \begin{align}
      \frac{a_2}{a_1}&=\exp{B_1\int \frac{dt}{V}},\label{4-2a}\\  
       \frac{a_3}{a_1}&=\exp{(B_1+B_2)\int \frac{dt}{V}},\label{4-2b}\\
        \frac{a_3}{a_2}&=\exp{B_2\int \frac{dt}{V}},\label{4-2c}
    \end{align}
\end{subequations}
where $\{B_1,B_2\}$ are constants of integration, and $V=a_1a_2a_3=a^3$ is the volume scale factor defined by (\ref{3-3}).
The straightforward solution of (\ref{4-1e}) is $a_1^2=a_2a_3$. substituting this solution into the definition of volume scale factor shows that the first directional scale factor equals the average scale factor, $a_1=a$. Also, if we insert solutions (\ref{4-2a})-(\ref{4-2c}) into relation $a_1^2=a_2a_3$, we find $B_2=-2B_1$. Therefore, the directional scale factors are not independent, and the set of  $\{a_1,a_2,a_3\}$ is reducible to the average scale factor
\begin{equation}
    \label{4-3}
   \begin{split}
       a_1(t)&=a(t),\\
       a_2(t)&=a(t)\exp{-B\int \frac{dt}{V}},\\
       a_3(t)&=a(t)\exp{B\int \frac{dt}{V}},
   \end{split} 
\end{equation}
where we redefined $B_1$ as $B$. Using Eqs. (\ref{4-3}) in Eqs. (\ref{4-1a})-(\ref{4-1c}) we obtain the analog of the Friedmann and the Raychaudhuri equations, respectively
\begin{subequations}
    \label{4-4}
    \begin{align}
& H^2-\frac{\beta^2}{a^2}=\frac{8\pi G_N}{3}(\rho+\rho^\text{(GDE)})+\frac{B^2}{3a^6},\label{4-4a}\\
 &\frac{\ddot a}{a}=-\frac{4\pi G_N}{3}(\rho+\rho^\text{(GDE)}+3p+3p^\text{(GDE)})-\frac{2B^2}{3a^6}.\label{4-4b}
    \end{align}
\end{subequations}

If the perfect fluid is assumed to satisfy the barotropic EoS $p=\omega\rho$, Eqs. (\ref{4-4a}) and (\ref{4-4b}) are reduced to
\begin{equation}\label{4-5}
    \begin{split}
        H^2&=\frac{8\pi G_N}{3}(\rho+\rho^\text{(GDE)})+\frac{\beta^2}{a^2}+\frac{B^2}{3a^6},\\
        \frac{\ddot a}{a}&=-\frac{4\pi G_N}{3}\Big\{(1+3\omega)\rho+(1+3\omega^\text{(GDE)})\rho^\text{(GDE)}\Big\}\\&-\frac{2B^2}{3a^6}.
    \end{split}
\end{equation}
Regarding $\nabla_\mu Q^{\mu\nu}=0$, which means 
\begin{equation}
    \frac{\dot \rho^\text{(GDE)}}{\rho^\text{(GDE)}}=-3(1+\omega_\text{(GDE)})H,
\end{equation}
 it can be verified
 \begin{equation}\label{4-6}
   \frac{\dot \rho}{\rho}=-3 (1+\omega)H-\frac{\dot G_N}{G_N}. 
 \end{equation}
This means that baryonic and dark matters are no longer conserved, and variations in the gravitational constant create (or annihilate) baryons and DM. However, according to Eq. (\ref{3-17}), the variation of $G_N$ is 
less than a 1\% during the Universe's 13.7 billion-year history. Therefore, the effect of $G_N$ in the local distribution of baryonic matter is insignificant. 
The modified conservation equation (\ref{4-6}) gives
\begin{equation}\label{4-7}
    G_N\rho(t)=G_0\rho_0a^{-3(1+\omega)},
\end{equation}
where $\rho_0$ is the matter density at the present epoch, and we normalized the current value of the mean scale factor, $a_0$, to unity. Using (\ref{3-22}) and (\ref{4-7}) in (\ref{4-5}) we obtain
\begin{multline}\label{4-8a}
            H^2=\frac{8\pi G_0}{3}\Big\{\rho_0a^{-3(1+\omega)}+\\
        \rho^\text{(GDE)}_0a^{-3(1+\omega_\text{(GDE)})}\Big\}+
        \frac{\beta^2}{a^2}+\frac{B^2}{3a^6},
\end{multline}       
and
\begin{multline}\label{4-8b}                \frac{\ddot a}{a}=-\frac{4\pi G_0}{3}\Big\{(1+3\omega)\rho_0a^{-3(1+\omega)}+\\(1+3\omega_\text{(GDE)})\rho^\text{(GDE)}_0a^{-3(1+\omega_\text{(GDE)})}\Big\}-\frac{2B^2}{3a^6},
\end{multline}
where $\omega_\text{(GDE)}=-\left(1+\frac{2\gamma}{3{\mathfrak N}}\right)$, and 
\begin{equation}
    \label{4-9}
    \rho^\text{(GDE)}_0=\frac{3{\mathfrak N}f_0^2}{8\pi G_0}=\frac{3{\mathfrak N}}{8\pi G_0L_0^2},
\end{equation}
is the density of the GDE at the present epoch, $t=t_0$.

It is simple to demonstrate that the expansions for the kinematic quantities $\sigma^2$,  the shear, and $\Theta$, the
expansion for the current model, are
\begin{equation}
    \begin{split}
        \Theta&=3H,\\
        \sigma^2&=\frac{B^2}{a^6}.
    \end{split}
\end{equation}
It is worth noting that for sufficiently large values of the average scale factor, i.e., $a(t)\rightarrow\infty$,
the ratio $\sigma\Theta$ vanishes. This implies that the Bianchi type-V space-time is asymptotically isotropic. However, the ratio of the Weyl curvature invariant, $C_{\mu\nu\alpha\beta}C^{\mu\nu\alpha\beta}$ to the square of the matter density, $\rho^2$, tends to zero only when
$\omega=1/3$. This suggests that only the type-V radiation-filled cosmologies can be asymptotically FLRW \cite{1986GReGr79N}. Additionally, unlike the pointlike singularity in FLRW cosmological models, the singularity of the type-V is cigar-shaped: when the singularity, $a\rightarrow0$, is approached, two of the scale factors $\{a_1,a_2,a_3\}$ in (\ref{4-3}) approach zero, while the third scale factor approaches infinity.

\section{Observational constraints}\label{Obser}
First, let us define the following definitions to have uniform physical parameters. Besides, the pressure of matter may be disregarded since, as we know, cold matter--both baryonic and dark--dominates the late-time Universe; therefore, $\omega=0$, and $\rho=\rho_0/a^3$. To begin, we define pressure and density for $\sigma^2$ and $\beta$, respectively 
\begin{eqnarray}\label{5-1}
    \begin{array}{ccc}
  p_\sigma=\rho_\sigma,&    \rho_\sigma=\rho_{0\sigma}a^{-6},&\rho_{0\sigma}=\frac{B^2}{8\pi G_0},\\
  p_\beta=-\frac{1}{3}\rho_\beta,& \rho_\beta=\rho_{0\beta}a^{-2},& \rho_{0\beta}=\frac{3\beta^2}{8\pi G_0}.
    \end{array}
\end{eqnarray}
Also, we define the density parameters of the GDE, $\Omega^\text{(GDE)}$, the cold matter $\Omega^\text{(m)}$, the shear scalar $\Omega^{(\sigma)}$, and $\Omega^{(\beta)}$ 
\begin{eqnarray}
    \label{5-2}
    \begin{array}{cc}
      \Omega^\text{(GDE)}_0=\frac{8\pi G_0\rho^\text{(GDE)}_0}{3H_0^2},&
      \Omega^\text{(m)}_0=\frac{8\pi G_0\rho_0}{3H_0^2},\\
      \Omega^{(\sigma)}_0=\frac{8\pi G_0\rho_{0\sigma}}{3H_0^2},& \Omega^{(\beta)}_0=\frac{8\pi G_0\rho_{0\beta}}{3H_0^2}.
    \end{array}
\end{eqnarray}
Here the index `0' indicates
that the variables are to be evaluated at the present time, $t=t_0$.
 The Friedmann and
the Raychaudhuri equations, (\ref{4-4a}) and (\ref{4-4b}) can be rewritten in terms of the redshift, $z$, density parameters, and  the deceleration parameter, $q=-\frac{\ddot a}{aH^2}$, as
\begin{multline}\label{5-3a}
 \left( \frac{H}{H_0}\right)^2=\Omega^\text{(m)}_0(1+z)^3+\Omega^\text{(GDE)}_0(1+z)^{3(\omega_K+1)}+\\\Omega^{(\sigma)}_0(1+z)^6+\Omega^{(\beta)}_0(1+z)^2,
  \end{multline}
  \begin{multline}\label{5-3b}  
2 q\left(\frac{H}{H_0}\right)^2=\Omega^\text{(m)}_0(1+z)^3+\\(1+3\omega_\text{(GDE)})\Omega^\text{(GDE)}_0(1+z)^{3(\omega_K+1)}+4\Omega^{(\sigma)}_0(1+z)^6. 
\end{multline}
 For present-day, Eq. (\ref{5-3a}) reduces to the  closure relation
 \begin{equation}
     \label{5-4}
\Omega_0^\text{(m)}+\Omega_0^\text{(GDE)}+\Omega_0^{(\sigma)}+\Omega_0^{(\beta)}=1.
 \end{equation}
As it is clear from Eqs. (\ref{5-3a}) and (\ref{5-3b}), they are identical to the  Friedmann and the Raychaudhuri equations in the standard FLRW cosmology with four perfect fluids: the cold matter $\omega_m=0$, the DE with a $\omega_\text{(GDE)}=-\left(1+\frac{2\gamma}{3{\mathfrak N}}\right)$, a $\sigma$-fluid with $\omega_\sigma=1$ which is identical to the stiff matter, and the spatial curvature `fluid' with $\omega_\beta=-1/3$.

 Also, the anisotropy parameter of the derived
model is
 \begin{multline}\label{5-5}
    A_p=
    {2\Omega^{(\sigma)}_0(1+z)^6}\Big\{\Omega^\text{(m)}_0(1+z)^3+\Omega^\text{(GDE)}_0(1+z)^{3(\omega_K+1)}\\+\Omega^{(\sigma)}_0(1+z)^6+\Omega^{(\beta)}_0(1+z)^2\Big\}^{-1}. 
 \end{multline}
 As redshift decreases, we notice that the anisotropy parameter $A_p$ reaches $A_p=2\Omega_0^{(\sigma)}$ at $z=0$. It suggests that the Universe initially had some anisotropy and eventually became almost isotropic. 
 
 The density parameter $\Omega^\text{(m)}_0$ is equal to the total of the baryon $\Omega^\text{(b)}_0$ and cold dark matter (CDM) $\Omega^\text{(CDM)}_0$ contributions, i.e. $\Omega^\text{(m)}=\Omega^\text{(CDM)}_0+\Omega^\text{(b)}_0$. The following equations  the WMAP 9-year (WMAP9) constraints on $\Omega^\text{(b)}_0$ and $\Omega^\text{(CDM)}_0$, respectively
 \begin{equation}
     \label{5-6}
     \begin{split}
\Omega^\text{(b)}_0h^2&=0.02264 \pm 0.00050,\\
\Omega^\text{(CDM)}_0h^2&=0.1138 \pm 0.0045.
\end{split}
\end{equation}
 If we take the value $h = 0.70$ \cite{2013ApJS19H}, then we have $\Omega^\text{(b)}_0=
0.0462041$, and $\Omega^{\text{(CDM)}}_0=0.232245$, for the central value. 
Thus,
\begin{equation}
    \label{matter}
    \Omega^\text{(m)}_0=0.278.
\end{equation}
The curvature of the Universe is constrained to be
close to the flat one ($\Omega^{(\beta)}_0 < 0.0085$) from WMAP9 data \cite{2013ApJS19H}. The density parameter of $\sigma$, $\Omega^{(\sigma)}_0$, is even less that the above value for $\Omega^{(\beta)}_0$. 
Also, according to WMAP9 data, the current density parameter of DE is
\begin{equation}\label{5-7}
    \Omega^\text{(DE)}_0=0.721 \pm 0.025.
\end{equation}
 
 The 95\% confidence level constraint on the EoS of the DE from WMAP9 data combined with additional observational data \cite{2013ApJS19H} is
 \begin{equation}\label{5-8}
 \omega^\text{(DE)}= \begin{cases}
  -1.084 \pm 0.063,~~~\beta=0,\\
 -1.122^{+0.068}_{-0.067} ~~~~~~~~~\beta\neq 0.
  \end{cases}
 \end{equation}
 Furthermore, using Hubble and Pantheon data from the late Universe ($z\leq 2.4$), the authors of Ref. \cite{2019Ph2A} get the constraint $\Omega^{(\sigma)}_0\leq 10^{-3}$, which is consistent with the model-independent requirements.

\section{Probing extra dimensions}\label{Prob} 
 
 As we saw in Eq. (\ref{4-9}), the energy density of the GDE at the present day is proportional to $1/L_0^2$. Thus, using Eq. (\ref{New11}), one can rewrite Eq. (\ref{4-9}) in therm of the thickness of the brane 
 \begin{multline}
    \label{6-1}
    \rho^\text{(GDE)}_0=\frac{3{\mathfrak N}}{8\pi G_0L_0^2}=\\
\frac{27{\mathfrak N}}{8d^2}\left(\frac{\Gamma(\frac{{\mathfrak N}+1}{2})}{\Gamma(\frac{{\mathfrak N}+2}{2})} \right)^2\left(\frac{e^2}{4\pi}\right)^{-2}\left(\frac{L_\text{Pl}}{l}\right)^6M_\text{Pl}^4  \Big|_{t=t_0}.
\end{multline}
To estimate $\rho^\text{(GDE)}_0$, we must first determine the thickness at the present time. In estimating the thickness, we must consider which part of the matter fields is less massive. The ordinary matter content of the Universe is Hadronic matter. Baryons are heavy and, according the Eq. (\ref{4-9}), their contribution can be neglected. On the other hand, muons are heavier than electrons and neutrinos but lighter than all other particles. Therefore, we set
\begin{equation}
    \label{6-2}
    l=\frac{\sqrt{{\mathfrak N}}\pi}{m_\mu},
\end{equation}
where $m_\mu=105.658$ MeV is the muon mass. Inserting the above equation into (\ref{6-1}), and considering $d=2$, we obtain
\begin{equation}\label{6-3}
   \rho^\text{(GDE)}_0=\frac{27}{32\pi^6{\mathfrak N}^2\alpha^2}\left(\frac{\Gamma(\frac{{\mathfrak N}+1}{2})}{\Gamma(\frac{{\mathfrak N}+2}{2})} \right)^2 \left(\frac{m_\mu}{M_\text{Pl}} \right)^6M_\text{Pl}^4\Big|_{t=t_0},
\end{equation}
where $\alpha=e^2/4\pi$ is the fine structure constant.

To fulfill today's cosmic acceleration, the energy density of DE must be on the order of the square of the current Hubble parameter, according to the standard model of cosmology. Let us go through this in further depth. Suppose that $\rho^\text{(DE)}_0$ is the current energy density of some sort of DE. According to the Friedmann equation, then we will have
\begin{equation}\label{6-4}
   \rho^\text{(DE)}_0=\frac{3H_0^2}{8\pi G_0}\Omega^\text{(DE)}_0=\frac{3}{8\pi}\left(\frac{H}{M_\text{Pl}} \right)^2\Omega^\text{(DE)}M_\text{Pl}^4\Big|_{t=t_0},
\end{equation}
 where $\Omega^\text{(DE)}_0$ is the density parameter of $\rho^\text{(DE)}$, and $M_\text{Pl}^4|_0$ is the Planck density at the present epoch. Regarding
 \begin{equation}
     \label{6-5}
     H_0=100 h ~\text{km}\cdot \text{sec}^{-1} \cdot\text{Mpc}^{-1} = 2.1332h\times 10^{-42}~ \text{GeV},
 \end{equation}
 where $h = 0.70 \pm 0.02$
\cite{2013ApJS19H}, and using (\ref{5-7}) we obtain
\begin{equation}
    \label{6-6}
\rho^\text{(DE)}_0=1.268\times10^{-123}M_\text{Pl}^4.    
\end{equation}

 In Table \ref{table1}, we find the energy density of the GDE (\ref{6-3}) for some specific number of extra
dimensions.
\begin{table}[h!]
    \centering
    \begin{tabular}{|c|c|}
    \hline
      ${\mathfrak N}$   & $\rho^\text{(GDE)}_0M_\text{Pl}^{-4}10^{123}$ \\
      \hline
      10&13.173\\
      \hline
      19&1.967\\
      \hline
      20& 1.688\\
      \hline
     21&1.460\\
      \hline
      \bf{ 22}&\bf{1.271}\\
      \hline
       23&1.114\\
      \hline
       24&0.981\\[1ex]
      \hline
          \end{tabular}
    \caption{The density parameter of the GDE obtained for
various number of extra dimensions ${\mathfrak N}$. {It should be noted that $\rho^\text{(GDE)}_0$ is computed using Eq. (\ref{6-3}).}}
    \label{table1}
\end{table}
It shows that for ${\mathfrak N}=22$, the energy density of GDE obtained in (\ref{6-3}) agrees with the observational DE density (\ref{6-6}).
  Note that, as is well known,  there is a significant discrepancy between the estimated Hubble constant values from Hubble Space Telescope (HST) observations of Cepheids in the Large Magellanic Cloud (LMC) \cite{Riess:2019cxk} and the Planck cosmic microwave background (CMB) \cite{Planck:2018vyg} observations. Consequently, this tension in $H_0$ is transmitted from the Hubble parameter measurement to the equation's computation of DE density. We can estimate the number of extra dimensions more accurately by more precise measurements of the density in Eq. (\ref{6-6}). It is interesting also to note that the average value of $\rho^\text{(DE)}_0=1.371\times10^{-123}M_\text{Pl}^4$ coincides with ${\mathfrak N}=22$ in Table \ref{table1}.  As we know, ${\mathfrak N}=22$ is the number of extra dimensions in the bosonic string theory, the original version of string theory. Due to the conformal anomaly, bosonic string theory exhibits contradictions in a generic space-time dimension. But, as Claud Lovelace initially observed \cite{Lovelace:1971fa,Cvetic:2011vz}, the anomaly cancels in a space-time of $26=1+3+22$ dimensions, which is the essential dimension for the theory.

Also, if we insert ${\mathfrak N}=22$ into Eq. (\ref{New11}), the current value of the normal curvature radii is
\begin{multline}
    \label{6-7}
    L_0=4.546\times 10^{61}L_\text{Pl}=\\7.348\times 10^{26}~\text{m}=23.812 ~\text{Gpc},
\end{multline} 
which is almost six times the present time Hubble radius $D_{H_0}=1/H_0=4.283$ Gpc (for $h=0.70$). Note that in the DGP brane model \cite{Deffayet:2001pu,Deffayet:2002sp} also, the effective size of extra dimension is in order of the present Hubble radius.

We can approach the problem differently. Starting with the equality of $\rho^\text{(GDE)}_0$ and $\rho^\text{(DE)}_0$ in Eqs. (\ref{6-3}) and (\ref{6-4}), we have the following equation
\begin{equation}
    \label{6-8}
    \Omega^\text{(DE)}={\mathfrak N}\left(\frac{D_{H_0}}{L_0} \right)^2=\frac{0.349}{h^2},
\end{equation}
 where in the second equality we used $D_{H_0}=1998/h$ Mpc, and ${\mathfrak N}=22$.
 The last equation gives $0.673<\Omega^\text{(DE)}_0<0.755$, which is consist with observations in (\ref{5-7}).

 Let us estimate the EoS of GDE, defined by Eq. (\ref{3-23}). 
 Regarding Eqs. (\ref{r1}), (\ref{3-15}), and (\ref{3-19}), the fine structure constant varies according to the following equation
 \begin{equation}\label{6-9}
     \alpha=\alpha_0(1+z)^{\gamma(1+\frac{1}{{\mathfrak N}})},
 \end{equation}
 where $z$ is the redshift, and $\alpha_0$
is the present value of the fine structure constant. 

Let us define the time variation of the fine structure constant at the recombination epoch relative to its present value
\begin{equation}
    \label{6-10}
    \frac{\Delta\alpha}{\alpha}=\frac{\alpha(t_\text{recom})-\alpha_0}{\alpha_0}.
\end{equation}
 Eq. (\ref{6-9}) gives us
 \begin{equation}
     \label{6-11}
  \frac{\Delta\alpha}{\alpha}=(1+z_\text{recom})^{\gamma(1+\frac{1}{{\mathfrak N}})}-1.   
 \end{equation}
 The 95\% confidence interval and the mean value of $\Delta\alpha/\alpha$ from WMAP9 data with
HST prior are \cite{2013ApJS19H}
 \begin{equation}\label{6-12}
 - 0.028 <\frac{\Delta\alpha}{\alpha} < 0.026,~~ \text{and}~~ \frac{\Delta\alpha}{\alpha}= -0.000894,     \end{equation}
 respectively. Using the mean value of $\Delta\alpha/\alpha$ in Eq. (\ref{6-11}) gives us
 \begin{equation}\label{6-13}
     \gamma=-1.22\times10^{-4}.
 \end{equation}
The above restriction on $\gamma$ and  Eq. (\ref{3-23}) show that the EoS of the GDE is equal to the EoS of the CC:
\begin{equation}\label{6-14}
    \omega_\text{(GDE)}=-(1+\frac{\gamma}{33})=-0.9999963=-1.
\end{equation}
This suggests that the time-variation of the fine structure and the gravitational constant, in the context of our model, are irrelevant throughout the Universe's history, with the exception of the very early Universe, when quantum gravity effects are crucial. Of course, our model is insufficient to capture Planck's area, and a quantum cosmology of the model is required. {Note that according to Eqs. (\ref{3-22}) and (\ref{6-13}) the energy density of GDE varies with time as
\begin{multline}
    \rho^{\text{(GDE)}}(t)=\\\rho^{\text{(GDE)}}_0\left(\frac{a}{a_0}\right)^{(\frac{2}{\mathfrak N}-1)\gamma}=\rho^{\text{(GDE)}}_0\left(\frac{a}{a_0}\right)^{1.11\times10^{-4}},
\end{multline}
where in the second equality, we assumed $\mathfrak N=22$, and we used Eq. (\ref{6-13}). 
The value of the scale factor at the end of inflation is $a(t_\text{inflation}=10^{-30}a_0$ \cite{Lineweaver:2003ie}. Inserting this value into the above equation, one can easily find the energy density of the GDE at the end of inflation
\begin{equation}
    \rho^{\text{(GDE)}}(t_\text{inflation})=0.992\rho^{\text{(GDE)}}_0.
\end{equation}
This suggests that the GDE density has not changed significantly throughout the Universe's history, and we can practically assume that $\gamma\simeq0$. It goes without saying that this model does not account for inflation, and our purpose here is to explore the late time Universe.}

The above value of the EoS parameter suggests that the GDE is the realization of the CC, $\Lambda$.
The equality $\rho^\text{(GDE)}=\Lambda/8\pi G_N$ gives us
\begin{equation}
    \label{6-15}
    \begin{split}
    \Lambda=\frac{3{\mathfrak N}}{L^2}&=\frac{27}{4\pi^5{\mathfrak N}^2}\left(\frac{\Gamma(\frac{{\mathfrak N}+1}{2})}{\Gamma(\frac{{\mathfrak N}+2}{2})}\right)^2\left(\frac{m_\mu}{M_\text{Pl}}\right)^6\frac{1}{\alpha^2L_\text{Pl}}\\
    &=1.222\times 10^{-52}~\frac{1}{\text{m}^2}.
    \end{split}
\end{equation}

\begin{figure}
  \centering
  \includegraphics[width=7cm]{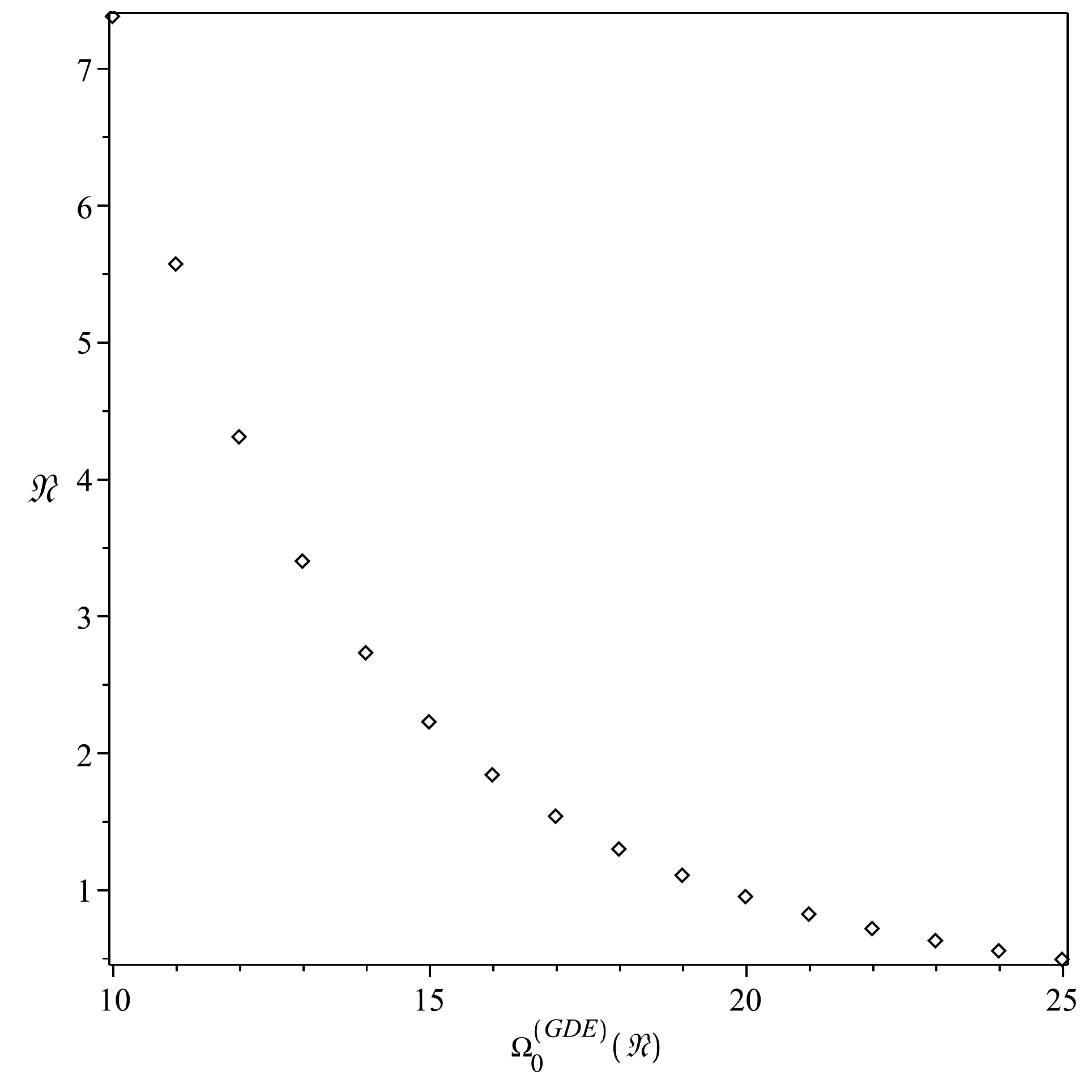}
  \caption{\small  The density parameter of the GDE at the present epoch, $\Omega^\text{(GDE)}_0$, vs. the number of extra dimensions ${\mathfrak N}$. We assumed $h=0.70$.}\label{fig2}
\end{figure}

 As we saw till now, the model requires twenty-two extra dimensions to be compatible with the experimental results, which raises an intriguing question. As we showed in the previous chapter, ten extra dimensions were adequate to provide a geometrical representation of the fundamental interactions. Thus, the question is why the model prefers the number twenty-two. As Fig. \ref{fig2} shows, the density parameter of the GDE increases rapidly for smaller values of the extra dimensions, in which its value reaches $\Omega^\text{(GDE)}_0>1$ for ${\mathfrak N}<20$. Hence, for ${\mathfrak N}<20$, the Universe enters the acceleration phase in the very early Universe, where the redshift is 
 \begin{equation}
     \label{redshift}
     z\simeq \frac{1}{\sqrt{2}}\left(\frac{\Omega^\text{(GDE)}_0}{\Omega^{(\sigma)}_0}\right)^\frac{1}{6}\gg1.
 \end{equation}
Consequently, matter condensations could not form in such a universe: Stars and galaxies will not be able to develop, and as a result, carbon will not create, and life as we know it will cease to exist.
Table \ref{table2} demonstrates the density parameter of the GDE at the present epoch, $\Omega^\text{(GDE)}_0$, for some selected number of the extra dimensions ${\mathfrak N}$. We assumed $h=0.70$.
As Table \ref{table2} shows, the DM content of the Universe for ${\mathfrak N}=20$ and ${\mathfrak N}=21$ is not enough, and the formation and evolution of galaxies and clusters do not happen, or as galaxies spin, they should be torn apart.
 \begin{table}[h!]
    \centering
    \begin{tabular}{|c|c|c|c|c|}
    \hline
      ${\mathfrak N}$   & $\Omega^\text{(DE)}$&$q_0$ & $z_t$  & $t_0$ (Gyr)\\
      \hline
      20& 0.945 & -0.918 & 2.261 & 20.44\\
      \hline
      21 & 0.818 & -0.727 & 1.078 & 15.40\\
      \hline
      \bf{22} & \bf{0.712} & \bf{-0.568} & \bf{0.704} & \bf{13.62}\\
      \hline
     23 & 0.624 & -0.436 & 0.491 & 12.63\\
      \hline
       24 & 0.549 & -0.324 & 0.346 & 11.98\\
      \hline
       25 & 0.487 & -0.23 & 0.237 & 11.52\\ [1ex]
           \hline
          \end{tabular}
    \caption{The density parameter of the GDE, the deceleration parameter at the present epoch, the deceleration to acceleration redshift $z_t$, and the age of the Universe obtained for the various number of extra dimensions ${\mathfrak N}$. To obtain these values, we used Eqs. (\ref{6-3}). }
    \label{table2}
\end{table}
On the other hand, for universes with more than twenty-two extra dimensions, the DM content is more than our Universe. As a result, these universes are younger, and the deceleration to the acceleration phase transition is more recent. 

As we know, direct calculation of the CC in effective quantum field theories yields an enormous discrepancy (exceeding 120 orders of magnitude) between the predicted value and the actual observed (and what we obtained above) density of the vacuum energy. In addition, the results are radiatively unstable. Let us clarify these points.
 
  In terms of the theory's locality and unitarity, the vacuum possesses energy in QFT. The vacuum loop diagrams for each matter field species must be determined to acquire the associated energy. Because they may always be absorbed into a redefinition of the subtraction scale, the finite contributions to vacuum energy are wholly arbitrary. Because QFT cannot theoretically anticipate the amount of the CC, we must measure it and suitably modify the finite component of $\rho_\text{vac}$ so that the theory meets cosmological data.

For example, suppose our scalar field is the standard model's Higgs boson. Given that observations constrain the total CC to $10^{-47}(\text{GeV})^4$: the limited contributions to vacuum energy at the 1-loop level must cancel to an accuracy of one part in $10^{47}$. After adjusting the bare CC to match observations at the 1-loop level, we must re-tune its value to a high degree of precision to cancel out the undesirable contributions at the 2-loop level. Similarly, when we proceed to 3-loops and beyond, its worth must be re-calibrated. In other words, it is radiatively unstable, and in loop perturbation theory, we must re-tune the bare CC in each order to deal with it. It is frequently referred to as the ``old'' CC problem since this radiatively unstable vacuum energy was an issue before the late time acceleration of the Universe was found. The ``new'' CC problem results from the finding that the observed Universe acceleration in late time is due to non-zero vacuum energy. The ``old'' CC issue is typically thought to have a solution that renders the vacuum energy precisely zero and radiatively stable. The reason why the observed CC is not exactly zero but rather a nonzero and extremely tiny number would then need to be explained. The CC problem is divided into two parts in the ``new'' version of the problem: First, Why is it so small? and Secondly, why do the energy densities of DE and the cold matter have the same order of magnitude at the present epoch? This is referred to as the coincidence problem.

In this regard, assuming that the vacuum energy estimated by QFT is precisely zero, Tables \ref{table1} and \ref{table2} illustrate a solution to the CC problems: Because we live in a Universe with ${\mathfrak N}=22$ non-compact extra dimensions, the CC is tiny, and its energy density is of the order of cold matter. {In other words, the Universe will speed up with the inclusion of 16 extra dimensions. The cosmos would accelerate sooner or later depending on whether there were more or fewer extra dimensions. }

\section{Zeldovich, Weinberg,  and Wesson's numerical relations}\label{Wave}
 
Let us recall some famous empirical formulas from the long history of the CC problem in effective field theories and gravity and encounter their relations to our findings in this paper.  Zeldovich was the first to construct a formula for the $\Lambda$ term from dimensional considerations \cite{Zeldovich:1967gd}, in which the CC represents the ground state energy of quantum fields, i.e., vacuum energy.
 He observed, in particular, that the leading proton contribution to zero-point energy (ZPE) is extremely huge compared to any cosmic density. After removing the extremely large leading term from the theoretical ZPE estimate, he realized that by returning to the proton's mass, $m_p$, as the typical mass scale of particle physics, he could obtain a much more reasonable order of magnitude estimate of the CC density using a second-order formula involving the natural presence of the gravitational constant. Then, he constructed the following formula using Eddington's and Dirac's ideas of large numbers and dimensional considerations
 \begin{equation}
     \label{7-1}
     \rho_\text{vac}\simeq G_Nm_p^6=\frac{m_p^6}{M_\text{Pl}^2}\simeq  10^{-38} ~\text{GeV}^4.
 \end{equation}
The aforementioned estimate is still nine orders of magnitude far from the measured value of the CC energy density. When the proton is replaced with the pion (whose mass is nearly ten times that of the proton), one gets a better estimate that differs from the observed value obtained in equation (\ref{6-6}) by just three orders of magnitude.
 
The above-mentioned method of changing the CC amount has a severe flaw. First and foremost, there is no hint in his argument regarding the radiative instability of the vacuum. Furthermore, there is no theoretical basis for determining the exact value of the CC.
As a result, the difficulty with the Zeldovich formula (\ref{7-1})  and its refinements is theoretical, and substituting a proton for a pion or other elementary particle will not address the problem.

On the other hand, we found in Eq. (\ref{6-3}) the following equation for vacuum energy 
\begin{equation}
    \label{7-2}
    \rho_0^\text{(GDE)}=C({\mathfrak N})\frac{G_N}{\alpha^2}m_\mu^6,
\end{equation}
where $C({\mathfrak N})$ is a function of the number of extra dimensions. There are similarities between this equation and Eq. (\ref{7-1}). However, instead of the pion of the proton, we have muon mass $m_\mu$, and it depends on the fine structure constant and the number of extra dimensions, ${\mathfrak N}$. QFT techniques cannot be used to obtain $C({\mathfrak N})$.

Weinberg offered a form of ``cosmic prediction'' of the pion mass in 1972 \cite{Weinberg:1972kfs} using a curious numerical equation
\begin{equation}
    \label{7-3}
m_\pi=\left(\frac{H_0}{G_N} \right)^\frac{1}{3},   
\end{equation}
where $m_\pi$ is the pion mass. and $H_0$ is the current value of the Hubble constant. One can obtain the equivalent relation of the above relation, using the relation $\Omega^\text{(GDE)}_0=\frac{8\pi G_0\rho^\text{(GDE)}_0}{3H_0^2}$ defined in (\ref{5-2}), $\Omega^\text{(GDE)}_0=\Omega^\text{(DE)}_0$, and (\ref{7-2})
\begin{equation}
    \label{7-4}
    m_\mu=\left(\frac{3\Omega^\text{(DE)}_0\alpha^2}{8\pi C({\mathfrak N})}\right)^\frac{1}{6}\left(\frac{ H_0}{G_N}\right)^\frac{1}{3},
\end{equation}
which gives us the exact value of the muon mass.

The ``quantum of mass,'' proposed by P. S. Wesson \cite{Wesson:2003qn}, is the last numerical-dimensional analysis connected to our work. Wesson derived two distinct masses using dimensional analysis, which can be derived from the set of constants $\{c, \hbar, G_N, \Lambda\}$ {of GR and quantum mechanics}. {It is worth noting that in this set of constants, $\Lambda$ and $G_N$ are the standard CC and the gravitational constant in $4D$ GR and do not include our induced GDE and $G_N$.} The relevant mass $M_\text{QS}$ at the ``quantum scale'' is
\begin{equation}
    \label{7-5}
    M_\text{QS}=\frac{\hbar}{c}\sqrt{\frac{\Lambda}{3}}=\sqrt{\frac{\Lambda}{3}},
\end{equation}
while the appropriate ``cosmic scale'' mass $M_\text{CS}$, according to Wesson, is
\begin{equation}
    \label{7-6}
    M_\text{CS}=\frac{c^2}{G_N}\sqrt{\frac{3}{\Lambda}}=\frac{1}{G_N}\sqrt{\frac{3}{\Lambda}}.
\end{equation}
The second mass, $M_\text{CS}$, has a simple interpretation: it is the mass of the observable Universe, which is equivalent to $10^{80}$ baryons of mass $10^{-21}$ Kg apiece. 

Comprehending the first mass, $M_\text{QS}$ is more challenging. 
To appreciate the relevance of ``quantum scale'' mass (\ref{7-5}) in our formalism, examine the Eq. (\ref{Gordon3}), and suppose a particle that is chargeless and can propagate conclusively to the extrinsic tube $\mathcal B_{\mathfrak N}$. In this case, as stated in Refs. \cite{Yazdani:2011ben,Jalalzadeh:2023liy}, Eqs. (\ref{Gordon2}) and (\ref{Gordon3}) represent the gravitational wave equations induced on the brane and orthogonal directions, respectively.

As we saw in section \ref{Geometry}, according to our assumption, all extra dimensions are spacelike, and $\mathcal B_{\mathfrak N}$ is locally
a ${\mathfrak N}$-sphere. In spherical coordinates, Eq. (\ref{Gordon2}) is
\begin{equation}
    \label{7-7}
  \bigg\{\frac{1}{L^2}\Delta_{S^{{\mathfrak N}}} +m^2\bigg\}\Phi_2(x^m)=0,  
\end{equation}
where $\Delta_{S^{{\mathfrak N}}}$ is the spherical Laplace operator in ${\mathfrak N}$-dimensions. 
Using the eigenvalue relation
\begin{equation}
    \label{7-8}
    \Delta_{S^{{\mathfrak N}-1}}Y_k=k(2-{\mathfrak N}-k)Y_k.
\end{equation}
in which $Y_k$, spherical harmonics, are the eigenvectors of the spherical Laplace operator \cite{Frye:2012jj}
we obtain
\begin{equation}
    \label{7-11}
    m_k=\frac{\sqrt{k+{\mathfrak N}-2}}{L}=\sqrt{1+\frac{k-2}{{\mathfrak N}}}\sqrt{\frac{\Lambda}{3}},
\end{equation}
where in the second equality of this equation, we used the relation between the {induced} CC and $L$ in Eq. (\ref{6-15}). 
 To obtain the $4D$ gravitational component of the gravitational wave equation, we need to obtain the components of the extrinsic potential, $\frac{1}{4}(\bar K_a\bar K^a-2\bar K_{\alpha\beta a}\bar K^{\alpha\beta a})$, in Eq. (\ref{Gordon2}). Inserting the components of the extrinsic curvature obtained in (\ref{3-11}) into the above potential,
{one can easily find
 \begin{equation}
     \frac{1}{4}(\bar K_a\bar K^a-2\bar K_{\alpha\beta a}\bar K^{\alpha\beta a})=\frac{3{\mathfrak N}}{L^2}=\Lambda,
 \end{equation}
where in obtaining the last equality, we used the definition of the induced $\Lambda$ in (\ref{6-15}).  As a result, 
  the wave equation (\ref{Gordon2}) simplifies into
\begin{multline}\label{GWave}
 \Big\{   -\frac{1}{\sqrt{|\bar g|}}\partial_\mu\bar g^{\mu\nu}\sqrt{|\bar g|}\partial_\nu +
 \Lambda+m^2\Big\}\Phi_1(x^\sigma)=0.
\end{multline}
Inserting the mass (\ref{7-11}) into the above equation gives us
\begin{multline}\label{GWave1}
 \Big\{   -\frac{1}{\sqrt{|\bar g|}}\partial_\mu\bar g^{\mu\nu}\sqrt{|\bar g|}\partial_\nu +4(1+\frac{k-2}{{4\mathfrak N}})\frac{\Lambda}{3}\Big\}\Phi_{1k}(x^\sigma)=0.
\end{multline}
Therefore, the zero mode ($k=0$, which is trapped in the brane, as needed) mass of the graviton at the presence of ${\mathfrak N}=22$ extra dimensions is
\begin{equation}
    \label{7-12}
    M_g=1.98\sqrt{\frac{\Lambda}{3}}=1.565\times 10^{-32}~\text{eV}.
\end{equation}
This demonstrates that Wesson's "quantum scale" mass is, in fact, the graviton's mass. It should be noted that the aforementioned value is compatible with various observational bounds on the graviton's mass; see, for example, Refs. \cite{Dvali:2002vf,Lue:2002sw,Choudhury:2002pu,Gruzinov:2001hp}. }

{Let us remind here of the remarkable finding made by Gazeau and Novello in Ref. \cite{Gazeau:2006uy}. They show that if the fundamental state of $4D$ space-time is Minkowskian, the graviton's square mass is proportional to the CC; otherwise if the fundamental state is de Sitter, the graviton is massless in the de Sitterian sense. Given that the bulk space in our work induces the CC (and, more broadly, GDE), the fundamental state in $4D$ brane is Minkowski, so Gazeau--Novello's hypothesis remains valid.}

{Moreover, R. Bousso contends in Ref. \cite{Bousso:2000nf} that the inverse of the CC serves as a limit for the total observable entropy. This is true for all space-times with a positive CC, including those with recollapsing universes and cosmologies dominated by ordinary matter \cite{Bousso:2000nf}.
Also, in \cite{Mongan:2006ay}, the author investigates a Friedmann universe with a vacuum dominating that is asymptotically approaching a de Sitter space and a cosmic event horizon whose area in Planck's units determines the maximum information that will ever be accessible to any observer. In particular, cosmological holography argues that the cosmological horizon of our universe contains all the information in it. Precisely, the holographic principle states that the total number of degrees of freedom, $\mathcal N$, (or entropy)  residing on the holographic screen is bounded to one-fourth of the area in Planck units \cite{Jalalzadeh:2017jdo}. Regarding that, the Universe is asymptotically a de Sitter, and utilizing the entropy of the de Sitter space \cite{201825E}
\begin{equation}
    S=\frac{3\pi}{G\Lambda}=\mathcal N,
\end{equation}
one can rewrite the mass of graviton in Eq. (\ref{7-12}) in terms of the total number of degrees of freedom, $\mathcal N$, and the Planck mass
\begin{equation}
    M_g\simeq\sqrt{\frac{4\pi}{\mathcal N}}M_\text{Pl}.
\end{equation}
Thus, we conclude that the total number of degrees of freedom living on the holographic screen may be described as the square of the ratio of the Planck mass to the graviton mass.}

\section{Conclusions}\label{Con}
If the CC is derived from particle physics' vacuum energy, it has a severe issue with its energy scale concerning DE density today \cite{Carroll:2000fy,Weinberg:1988cp}. The hope that quantum gravity will explain why the vacuum does not gravitate is a plausible approach to the first part of the CC problem \cite{ellismaarte}. As a result, one may claim that the CC, or DE, has a gravitational-geometrical basis and is unrelated to vacuum energy.

Based on Refs. \cite{Jalalzadeh:2013wza,Rostami:2015ixa}, we developed a novel gravitational model in which our space-time is a braneworld immersed in a $26D$ space-time to address the old and new CC problems and find the origin of geometrical vacuum energy density. We show that one can conceive a brane model with thickness using the Nash--Morse implicit function theorem.

We applied the model for a Bianchi type-V cosmological model. The gravitational constant is shown to be a function of the extrinsic normal curvature of space-time, the thickness of the Universe along extra dimensions, the fine structure constant, and the number of extra dimensions. The model predicts DE density in terms of the aforementioned fundamental constants. Our model predicts the following values for the energy density of the DE at present $\Omega^\text{DE)}_0=0.712$, for the deceleration parameter at the present epoch $q_0=-0.568$, the redshift at the transition epoch $z_t=0.704$, and the age of the Universe $t_0=13.62$ Gyr. Calculated values are compatible with observations. These results are highly sensitive to the number of extra dimensions. If we include more or less than 22 extra dimensions in the model equations, the results will differ dramatically from the experimental observations. Therefore, the model suggests a simple solution for the coincidence problem: $\rho_0^{(\Lambda)}$ is of the order
of the present matter density $\rho^\text{(m)}(t_0)$ because the Universe has $\mathfrak{N}=22$ non-compact extra dimensions.

{Note that this does not mean that the GDE only applies to space with an extra 22 dimensions (or any other specific number). The model works for a sufficiently higher number of extra dimensions. For example, for 25 extra dimensions, the Universe must be younger at the present epoch, and the transition to the acceleration phase would have occurred much closer to the present. Also, the numerals in the first and second tables are (adequately) approximations for our model. Nevertheless, if we want to be more precise, the model's parameters should be accurately fitted with observational data. Here our goal is to show that there is a consistent and reasonable solution, using available empirical data, for the case of evaluating the number of extra dimensions in the Universe.}

Furthermore, we show that the numerical equations of Zeldovich, Weinberg, and Wesson have straightforward interpretations. {The Wesson's mass is reinterpreted as the graviton's mass. In terms of the holographic principle, we infer that the total number of degrees of freedom living on the holographic screen may be defined as the square of the Planck mass to graviton mass ratio.}

\section*{CRediT authorship contribution statement}
R. Jalalzadeh: Investigation, Writing--original draft, Software, Formal analysis, Visualization.  S. Jalalzadeh: Conceptualization,
Writing -- review \& editing, Validation, Supervision. B. Malekolkalami: Writing --review
\& editing, Validation, Visualization.

\section*{Declaration of competing interest}
The authors declare that he has no known competing financial interests or personal relationships that could have appeared to influence the work reported in this paper.

\section*{Data availability}
No data was used for the research described in the article.

\section*{Acknowledgements}
We thank the anonymous Referees for their time and work reviewing the manuscript. We sincerely appreciate all helpful comments and recommendations that helped us improve the manuscript's quality. 

S.J. acknowledges financial support from the National Council for Scientific and Technological Development--CNPq, Grant no. 308131/2022-3. 
  
\bibliographystyle{elsarticle-num}

\bibliography{main}

\begin{thebibliography}{100}
\expandafter\ifx\csname url\endcsname\relax
  \def\url#1{\texttt{#1}}\fi
\expandafter\ifx\csname urlprefix\endcsname\relax\def\urlprefix{URL }\fi
\expandafter\ifx\csname href\endcsname\relax
  \def\href#1#2{#2} \def\path#1{#1}\fi

\bibitem{2007p2221N}
G.~{Nordstr{\"o}m}, {On the possibility of unifying the electromagnetic and the
  gravitational fields}, arXiv e-prints (2007) physics/0702221\href
  {http://arxiv.org/abs/physics/0702221} {\path{arXiv:physics/0702221}}, \href
  {http://dx.doi.org/10.48550/arXiv.physics/0702221}
  {\path{doi:10.48550/arXiv.physics/0702221}}.

\bibitem{Kaluza:1921tu}
T.~Kaluza, {Zum Unit\"atsproblem der Physik}, Sitzungsber. Preuss. Akad. Wiss.
  Berlin (Math. Phys. ) 1921 (1921) 966--972.
\newblock \href {http://arxiv.org/abs/1803.08616} {\path{arXiv:1803.08616}},
  \href {http://dx.doi.org/10.1142/S0218271818700017}
  {\path{doi:10.1142/S0218271818700017}}.

\bibitem{1926ZPhy95K}
O.~{Klein}, {Quantentheorie und f{\"u}nfdimensionale Relativit{\"a}tstheorie},
  Zeitschrift fur Physik 37~(12) (1926) 895--906.
\newblock \href {http://dx.doi.org/10.1007/BF01397481}
  {\path{doi:10.1007/BF01397481}}.

\bibitem{ellismaarte}
{Ellis, George F. R. and Maartens, Roy and MacCallum, Malcolm A. H.},
  {Relativistic Cosmology}, {Cambridge University Press}, 2012.
\newblock \href {http://dx.doi.org/10.1017/CBO9781139014403}
  {\path{doi:10.1017/CBO9781139014403}}.

\bibitem{DeFelice:2010aj}
A.~De~Felice, S.~Tsujikawa, {f(R) theories}, Living Rev. Rel. 13 (2010) 3.
\newblock \href {http://arxiv.org/abs/1002.4928} {\path{arXiv:1002.4928}},
  \href {http://dx.doi.org/10.12942/lrr-2010-3}
  {\path{doi:10.12942/lrr-2010-3}}.

\bibitem{Nojiri:2010wj}
S.~Nojiri, S.~D. Odintsov, {Unified cosmic history in modified gravity: from
  F(R) theory to Lorentz non-invariant models}, Phys. Rept. 505 (2011) 59--144.
\newblock \href {http://arxiv.org/abs/1011.0544} {\path{arXiv:1011.0544}},
  \href {http://dx.doi.org/10.1016/j.physrep.2011.04.001}
  {\path{doi:10.1016/j.physrep.2011.04.001}}.

\bibitem{Clifton:2011jh}
T.~Clifton, P.~G. Ferreira, A.~Padilla, C.~Skordis, {Modified Gravity and
  Cosmology}, Phys. Rept. 513 (2012) 1--189.
\newblock \href {http://arxiv.org/abs/1106.2476} {\path{arXiv:1106.2476}},
  \href {http://dx.doi.org/10.1016/j.physrep.2012.01.001}
  {\path{doi:10.1016/j.physrep.2012.01.001}}.

\bibitem{Miranda:2009rs}
V.~Miranda, S.~E. Joras, I.~Waga, M.~Quartin, {Viable Singularity-Free f(R)
  Gravity Without a Cosmological Constant}, Phys. Rev. Lett. 102 (2009) 221101.
\newblock \href {http://arxiv.org/abs/0905.1941} {\path{arXiv:0905.1941}},
  \href {http://dx.doi.org/10.1103/PhysRevLett.102.221101}
  {\path{doi:10.1103/PhysRevLett.102.221101}}.

\bibitem{Amendola:2006we}
L.~Amendola, R.~Gannouji, D.~Polarski, S.~Tsujikawa, {Conditions for the
  cosmological viability of f(R) dark energy models}, Phys. Rev. D 75 (2007)
  083504.
\newblock \href {http://arxiv.org/abs/gr-qc/0612180}
  {\path{arXiv:gr-qc/0612180}}, \href
  {http://dx.doi.org/10.1103/PhysRevD.75.083504}
  {\path{doi:10.1103/PhysRevD.75.083504}}.

\bibitem{Amendola:1999qq}
L.~Amendola, {Scaling solutions in general nonminimal coupling theories}, Phys.
  Rev. D 60 (1999) 043501.
\newblock \href {http://arxiv.org/abs/astro-ph/9904120}
  {\path{arXiv:astro-ph/9904120}}, \href
  {http://dx.doi.org/10.1103/PhysRevD.60.043501}
  {\path{doi:10.1103/PhysRevD.60.043501}}.

\bibitem{Uzan:1999ch}
J.-P. Uzan, {Cosmological scaling solutions of nonminimally coupled scalar
  fields}, Phys. Rev. D 59 (1999) 123510.
\newblock \href {http://arxiv.org/abs/gr-qc/9903004}
  {\path{arXiv:gr-qc/9903004}}, \href
  {http://dx.doi.org/10.1103/PhysRevD.59.123510}
  {\path{doi:10.1103/PhysRevD.59.123510}}.

\bibitem{Bartolo:1999sq}
N.~Bartolo, M.~Pietroni, {Scalar tensor gravity and quintessence}, Phys. Rev. D
  61 (2000) 023518.
\newblock \href {http://arxiv.org/abs/hep-ph/9908521}
  {\path{arXiv:hep-ph/9908521}}, \href
  {http://dx.doi.org/10.1103/PhysRevD.61.023518}
  {\path{doi:10.1103/PhysRevD.61.023518}}.

\bibitem{Esposito-Farese:2000pbo}
G.~Esposito-Farese, D.~Polarski, {Scalar tensor gravity in an accelerating
  universe}, Phys. Rev. D 63 (2001) 063504.
\newblock \href {http://arxiv.org/abs/gr-qc/0009034}
  {\path{arXiv:gr-qc/0009034}}, \href
  {http://dx.doi.org/10.1103/PhysRevD.63.063504}
  {\path{doi:10.1103/PhysRevD.63.063504}}.

\bibitem{Koivisto:2006xf}
T.~Koivisto, D.~F. Mota, {Cosmology and Astrophysical Constraints of
  Gauss-Bonnet Dark Energy}, Phys. Lett. B 644 (2007) 104--108.
\newblock \href {http://arxiv.org/abs/astro-ph/0606078}
  {\path{arXiv:astro-ph/0606078}}, \href
  {http://dx.doi.org/10.1016/j.physletb.2006.11.048}
  {\path{doi:10.1016/j.physletb.2006.11.048}}.

\bibitem{Koivisto:2006ai}
T.~Koivisto, D.~F. Mota, {Gauss-Bonnet Quintessence: Background Evolution,
  Large Scale Structure and Cosmological Constraints}, Phys. Rev. D 75 (2007)
  023518.
\newblock \href {http://arxiv.org/abs/hep-th/0609155}
  {\path{arXiv:hep-th/0609155}}, \href
  {http://dx.doi.org/10.1103/PhysRevD.75.023518}
  {\path{doi:10.1103/PhysRevD.75.023518}}.

\bibitem{Tsujikawa:2006ph}
S.~Tsujikawa, M.~Sami, {String-inspired cosmology: Late time transition from
  scaling matter era to dark energy universe caused by a Gauss-Bonnet
  coupling}, JCAP 01 (2007) 006.
\newblock \href {http://arxiv.org/abs/hep-th/0608178}
  {\path{arXiv:hep-th/0608178}}, \href
  {http://dx.doi.org/10.1088/1475-7516/2007/01/006}
  {\path{doi:10.1088/1475-7516/2007/01/006}}.

\bibitem{Fujii:1982ms}
Y.~Fujii, {Origin of the Gravitational Constant and Particle Masses in Scale
  Invariant Scalar - Tensor Theory}, Phys. Rev. D 26 (1982) 2580.
\newblock \href {http://dx.doi.org/10.1103/PhysRevD.26.2580}
  {\path{doi:10.1103/PhysRevD.26.2580}}.

\bibitem{Ford:1987de}
L.~H. Ford, {COSMOLOGICAL CONSTANT DAMPING BY UNSTABLE SCALAR FIELDS}, Phys.
  Rev. D 35 (1987) 2339.
\newblock \href {http://dx.doi.org/10.1103/PhysRevD.35.2339}
  {\path{doi:10.1103/PhysRevD.35.2339}}.

\bibitem{Wetterich:1987fm}
C.~Wetterich, {Cosmology and the Fate of Dilatation Symmetry}, Nucl. Phys. B
  302 (1988) 668--696.
\newblock \href {http://arxiv.org/abs/1711.03844} {\path{arXiv:1711.03844}},
  \href {http://dx.doi.org/10.1016/0550-3213(88)90193-9}
  {\path{doi:10.1016/0550-3213(88)90193-9}}.

\bibitem{Ratra:1987rm}
B.~Ratra, P.~J.~E. Peebles, {Cosmological Consequences of a Rolling Homogeneous
  Scalar Field}, Phys. Rev. D 37 (1988) 3406.
\newblock \href {http://dx.doi.org/10.1103/PhysRevD.37.3406}
  {\path{doi:10.1103/PhysRevD.37.3406}}.

\bibitem{Chiba:1999ka}
T.~Chiba, T.~Okabe, M.~Yamaguchi, {Kinetically driven quintessence}, Phys. Rev.
  D 62 (2000) 023511.
\newblock \href {http://arxiv.org/abs/astro-ph/9912463}
  {\path{arXiv:astro-ph/9912463}}, \href
  {http://dx.doi.org/10.1103/PhysRevD.62.023511}
  {\path{doi:10.1103/PhysRevD.62.023511}}.

\bibitem{Armendariz-Picon:2000nqq}
C.~Armendariz-Picon, V.~F. Mukhanov, P.~J. Steinhardt, {A Dynamical solution to
  the problem of a small cosmological constant and late time cosmic
  acceleration}, Phys. Rev. Lett. 85 (2000) 4438--4441.
\newblock \href {http://arxiv.org/abs/astro-ph/0004134}
  {\path{arXiv:astro-ph/0004134}}, \href
  {http://dx.doi.org/10.1103/PhysRevLett.85.4438}
  {\path{doi:10.1103/PhysRevLett.85.4438}}.

\bibitem{Armendariz-Picon:2000ulo}
C.~Armendariz-Picon, V.~F. Mukhanov, P.~J. Steinhardt, {Essentials of k
  essence}, Phys. Rev. D 63 (2001) 103510.
\newblock \href {http://arxiv.org/abs/astro-ph/0006373}
  {\path{arXiv:astro-ph/0006373}}, \href
  {http://dx.doi.org/10.1103/PhysRevD.63.103510}
  {\path{doi:10.1103/PhysRevD.63.103510}}.

\bibitem{Singh:2003vx}
P.~Singh, M.~Sami, N.~Dadhich, {Cosmological dynamics of phantom field}, Phys.
  Rev. D 68 (2003) 023522.
\newblock \href {http://arxiv.org/abs/hep-th/0305110}
  {\path{arXiv:hep-th/0305110}}, \href
  {http://dx.doi.org/10.1103/PhysRevD.68.023522}
  {\path{doi:10.1103/PhysRevD.68.023522}}.

\bibitem{Sami:2003xv}
M.~Sami, A.~Toporensky, {Phantom field and the fate of universe}, Mod. Phys.
  Lett. A 19 (2004) 1509.
\newblock \href {http://arxiv.org/abs/gr-qc/0312009}
  {\path{arXiv:gr-qc/0312009}}, \href
  {http://dx.doi.org/10.1142/S0217732304013921}
  {\path{doi:10.1142/S0217732304013921}}.

\bibitem{Carroll:2003st}
S.~M. Carroll, M.~Hoffman, M.~Trodden, {Can the dark energy equation-of-state
  parameter $w$ be less than $-1$?}, Phys. Rev. D 68 (2003) 023509.
\newblock \href {http://arxiv.org/abs/astro-ph/0301273}
  {\path{arXiv:astro-ph/0301273}}, \href
  {http://dx.doi.org/10.1103/PhysRevD.68.023509}
  {\path{doi:10.1103/PhysRevD.68.023509}}.

\bibitem{Amendola:1999er}
L.~Amendola, {Coupled quintessence}, Phys. Rev. D 62 (2000) 043511.
\newblock \href {http://arxiv.org/abs/astro-ph/9908023}
  {\path{arXiv:astro-ph/9908023}}, \href
  {http://dx.doi.org/10.1103/PhysRevD.62.043511}
  {\path{doi:10.1103/PhysRevD.62.043511}}.

\bibitem{Dalal:2001dt}
N.~Dalal, K.~Abazajian, E.~E. Jenkins, A.~V. Manohar, {Testing the cosmic
  coincidence problem and the nature of dark energy}, Phys. Rev. Lett. 87
  (2001) 141302.
\newblock \href {http://arxiv.org/abs/astro-ph/0105317}
  {\path{arXiv:astro-ph/0105317}}, \href
  {http://dx.doi.org/10.1103/PhysRevLett.87.141302}
  {\path{doi:10.1103/PhysRevLett.87.141302}}.

\bibitem{Zimdahl:2001ar}
W.~Zimdahl, D.~Pavon, {Interacting quintessence}, Phys. Lett. B 521 (2001)
  133--138.
\newblock \href {http://arxiv.org/abs/astro-ph/0105479}
  {\path{arXiv:astro-ph/0105479}}, \href
  {http://dx.doi.org/10.1016/S0370-2693(01)01174-1}
  {\path{doi:10.1016/S0370-2693(01)01174-1}}.

\bibitem{Bertacca:2007ux}
D.~Bertacca, S.~Matarrese, M.~Pietroni, {Unified Dark Matter in Scalar Field
  Cosmologies}, Mod. Phys. Lett. A 22 (2007) 2893--2907.
\newblock \href {http://arxiv.org/abs/astro-ph/0703259}
  {\path{arXiv:astro-ph/0703259}}, \href
  {http://dx.doi.org/10.1142/S0217732307025893}
  {\path{doi:10.1142/S0217732307025893}}.

\bibitem{Fukuyama:2007sx}
T.~Fukuyama, M.~Morikawa, T.~Tatekawa, {Cosmic structures via Bose Einstein
  condensation and its collapse}, JCAP 06 (2008) 033.
\newblock \href {http://arxiv.org/abs/0705.3091} {\path{arXiv:0705.3091}},
  \href {http://dx.doi.org/10.1088/1475-7516/2008/06/033}
  {\path{doi:10.1088/1475-7516/2008/06/033}}.

\bibitem{Tomita:1999qn}
K.~Tomita, {Distances and lensing in cosmological void models}, Astrophys. J.
  529 (2000) 38.
\newblock \href {http://arxiv.org/abs/astro-ph/9906027}
  {\path{arXiv:astro-ph/9906027}}, \href {http://dx.doi.org/10.1086/308277}
  {\path{doi:10.1086/308277}}.

\bibitem{Tomita:2000jj}
K.~Tomita, {A local void and the accelerating universe}, Mon. Not. Roy. Astron.
  Soc. 326 (2001) 287.
\newblock \href {http://arxiv.org/abs/astro-ph/0011484}
  {\path{arXiv:astro-ph/0011484}}, \href
  {http://dx.doi.org/10.1046/j.1365-8711.2001.04597.x}
  {\path{doi:10.1046/j.1365-8711.2001.04597.x}}.

\bibitem{Celerier:1999hp}
M.-N. Celerier, {Do we really see a cosmological constant in the supernovae
  data?}, Astron. Astrophys. 353 (2000) 63--71.
\newblock \href {http://arxiv.org/abs/astro-ph/9907206}
  {\path{arXiv:astro-ph/9907206}}.

\bibitem{Alnes:2005rw}
H.~Alnes, M.~Amarzguioui, O.~Gron, {An inhomogeneous alternative to dark
  energy?}, Phys. Rev. D 73 (2006) 083519.
\newblock \href {http://arxiv.org/abs/astro-ph/0512006}
  {\path{arXiv:astro-ph/0512006}}, \href
  {http://dx.doi.org/10.1103/PhysRevD.73.083519}
  {\path{doi:10.1103/PhysRevD.73.083519}}.

\bibitem{19865O}
F.~{Occhionero}, P.~{Santangelo}, N.~{Vittorio}, {Holes in cosmology}, Astron.
  Astrophys. 117~(2) (1983) 365--367.

\bibitem{Hashemi:2014lqa}
S.~S. Hashemi, S.~Jalalzadeh, N.~Riazi, {Dark side of the universe in the
  Stephani cosmology}, Eur. Phys. J. C 74~(8) (2014) 2995.
\newblock \href {http://arxiv.org/abs/1401.2429} {\path{arXiv:1401.2429}},
  \href {http://dx.doi.org/10.1140/epjc/s10052-014-2995-z}
  {\path{doi:10.1140/epjc/s10052-014-2995-z}}.

\bibitem{Rasanen:2003fy}
S.~Rasanen, {Dark energy from backreaction}, JCAP 02 (2004) 003.
\newblock \href {http://arxiv.org/abs/astro-ph/0311257}
  {\path{arXiv:astro-ph/0311257}}, \href
  {http://dx.doi.org/10.1088/1475-7516/2004/02/003}
  {\path{doi:10.1088/1475-7516/2004/02/003}}.

\bibitem{Kolb:2005da}
E.~W. Kolb, S.~Matarrese, A.~Riotto, {On cosmic acceleration without dark
  energy}, New J. Phys. 8 (2006) 322.
\newblock \href {http://arxiv.org/abs/astro-ph/0506534}
  {\path{arXiv:astro-ph/0506534}}, \href
  {http://dx.doi.org/10.1088/1367-2630/8/12/322}
  {\path{doi:10.1088/1367-2630/8/12/322}}.

\bibitem{Hirata:2005ei}
C.~M. Hirata, U.~Seljak, {Can superhorizon cosmological perturbations explain
  the acceleration of the Universe?}, Phys. Rev. D 72 (2005) 083501.
\newblock \href {http://arxiv.org/abs/astro-ph/0503582}
  {\path{arXiv:astro-ph/0503582}}, \href
  {http://dx.doi.org/10.1103/PhysRevD.72.083501}
  {\path{doi:10.1103/PhysRevD.72.083501}}.

\bibitem{Waeming:2021ytf}
A.~Waeming, T.~Klangburam, C.~Pongkitivanichkul, D.~Samart, {Dark matter and
  dark energy from a Kaluza\textendash{}Klein inspired Brans\textendash{}Dicke
  gravity with barotropic fluid}, Eur. Phys. J. C 82~(5) (2022) 409.
\newblock \href {http://arxiv.org/abs/2107.00678} {\path{arXiv:2107.00678}},
  \href {http://dx.doi.org/10.1140/epjc/s10052-022-10355-4}
  {\path{doi:10.1140/epjc/s10052-022-10355-4}}.

\bibitem{Pongkitivanichkul:2020txi}
C.~Pongkitivanichkul, D.~Samart, N.~Thongyoi, N.~Lunrasri, {A
  Kaluza\textendash{}Klein inspired Brans\textendash{}Dicke gravity with dark
  matter and dark energy model}, Phys. Dark Univ. 30 (2020) 100731.
\newblock \href {http://arxiv.org/abs/2005.08791} {\path{arXiv:2005.08791}},
  \href {http://dx.doi.org/10.1016/j.dark.2020.100731}
  {\path{doi:10.1016/j.dark.2020.100731}}.

\bibitem{PoncedeLeon:2010kh}
J.~Ponce~de Leon, {Brans-Dicke Cosmology in 4D from scalar-vacuum in 5D}, JCAP
  03 (2010) 030.
\newblock \href {http://arxiv.org/abs/1001.1961} {\path{arXiv:1001.1961}},
  \href {http://dx.doi.org/10.1088/1475-7516/2010/03/030}
  {\path{doi:10.1088/1475-7516/2010/03/030}}.

\bibitem{Chen:2009ep}
S.~Chen, J.~Jing, {The torsion cosmology in Kaluza-Klein theory}, JCAP 09
  (2009) 001.
\newblock \href {http://arxiv.org/abs/0905.0302} {\path{arXiv:0905.0302}},
  \href {http://dx.doi.org/10.1088/1475-7516/2009/09/001}
  {\path{doi:10.1088/1475-7516/2009/09/001}}.

\bibitem{Pietroni:2002ey}
M.~Pietroni, {Brane worlds and the cosmic coincidence problem}, Phys. Rev. D 67
  (2003) 103523.
\newblock \href {http://arxiv.org/abs/hep-ph/0203085}
  {\path{arXiv:hep-ph/0203085}}, \href
  {http://dx.doi.org/10.1103/PhysRevD.67.103523}
  {\path{doi:10.1103/PhysRevD.67.103523}}.

\bibitem{Antoniadis:1998ig}
I.~Antoniadis, N.~Arkani-Hamed, S.~Dimopoulos, G.~R. Dvali, {New dimensions at
  a millimeter to a Fermi and superstrings at a TeV}, Phys. Lett. B 436 (1998)
  257--263.
\newblock \href {http://arxiv.org/abs/hep-ph/9804398}
  {\path{arXiv:hep-ph/9804398}}, \href
  {http://dx.doi.org/10.1016/S0370-2693(98)00860-0}
  {\path{doi:10.1016/S0370-2693(98)00860-0}}.

\bibitem{Johannsen:2008aa}
T.~Johannsen, {Constraints on the Size of Extra Dimensions from the Orbital
  Evolution of the Black-Hole X-Ray Binary XTE J1118+480}, Astron. Astrophys.
  507 (2009) 617.
\newblock \href {http://arxiv.org/abs/0812.0809} {\path{arXiv:0812.0809}},
  \href {http://dx.doi.org/10.1051/0004-6361/200912803}
  {\path{doi:10.1051/0004-6361/200912803}}.

\bibitem{Salumbides:2015qwa}
E.~J. Salumbides, A.~N. Schellekens, B.~Gato-Rivera, W.~Ubachs, {Constraints on
  extra dimensions from precision molecular spectroscopy}, New J. Phys. 17~(3)
  (2015) 033015.
\newblock \href {http://arxiv.org/abs/1502.02838} {\path{arXiv:1502.02838}},
  \href {http://dx.doi.org/10.1088/1367-2630/17/3/033015}
  {\path{doi:10.1088/1367-2630/17/3/033015}}.

\bibitem{Chakravarti:2019aup}
K.~Chakravarti, S.~Chakraborty, K.~S. Phukon, S.~Bose, S.~SenGupta,
  {Constraining extra-spatial dimensions with observations of GW170817}, Class.
  Quant. Grav. 37~(10) (2020) 105004.
\newblock \href {http://arxiv.org/abs/1903.10159} {\path{arXiv:1903.10159}},
  \href {http://dx.doi.org/10.1088/1361-6382/ab8355}
  {\path{doi:10.1088/1361-6382/ab8355}}.

\bibitem{Pardo:2018ipy}
K.~Pardo, M.~Fishbach, D.~E. Holz, D.~N. Spergel, {Limits on the number of
  spacetime dimensions from GW170817}, JCAP 07 (2018) 048.
\newblock \href {http://arxiv.org/abs/1801.08160} {\path{arXiv:1801.08160}},
  \href {http://dx.doi.org/10.1088/1475-7516/2018/07/048}
  {\path{doi:10.1088/1475-7516/2018/07/048}}.

\bibitem{Vagnozzi:2019apd}
S.~Vagnozzi, L.~Visinelli, {Hunting for extra dimensions in the shadow of
  M87*}, Phys. Rev. D 100~(2) (2019) 024020.
\newblock \href {http://arxiv.org/abs/1905.12421} {\path{arXiv:1905.12421}},
  \href {http://dx.doi.org/10.1103/PhysRevD.100.024020}
  {\path{doi:10.1103/PhysRevD.100.024020}}.

\bibitem{Visinelli:2017bny}
L.~Visinelli, N.~Bolis, S.~Vagnozzi, {Brane-world extra dimensions in light of
  GW170817}, Phys. Rev. D 97~(6) (2018) 064039.
\newblock \href {http://arxiv.org/abs/1711.06628} {\path{arXiv:1711.06628}},
  \href {http://dx.doi.org/10.1103/PhysRevD.97.064039}
  {\path{doi:10.1103/PhysRevD.97.064039}}.

\bibitem{Corman:2020pyr}
M.~Corman, C.~Escamilla-Rivera, M.~A. Hendry, {Constraining extra dimensions on
  cosmological scales with LISA future gravitational wave siren data}, JCAP 02
  (2021) 005.
\newblock \href {http://arxiv.org/abs/2004.04009} {\path{arXiv:2004.04009}},
  \href {http://dx.doi.org/10.1088/1475-7516/2021/02/005}
  {\path{doi:10.1088/1475-7516/2021/02/005}}.

\bibitem{Corman:2021avn}
M.~Corman, A.~Ghosh, C.~Escamilla-Rivera, M.~A. Hendry, S.~Marsat, N.~Tamanini,
  {Constraining cosmological extra dimensions with gravitational wave standard
  sirens: From theory to current and future multimessenger observations}, Phys.
  Rev. D 105~(6) (2022) 064061.
\newblock \href {http://arxiv.org/abs/2109.08748} {\path{arXiv:2109.08748}},
  \href {http://dx.doi.org/10.1103/PhysRevD.105.064061}
  {\path{doi:10.1103/PhysRevD.105.064061}}.

\bibitem{Du:2020rlx}
Y.~Du, S.~Tahura, D.~Vaman, K.~Yagi, {Probing Compactified Extra Dimensions
  with Gravitational Waves}, Phys. Rev. D 103~(4) (2021) 044031.
\newblock \href {http://arxiv.org/abs/2004.03051} {\path{arXiv:2004.03051}},
  \href {http://dx.doi.org/10.1103/PhysRevD.103.044031}
  {\path{doi:10.1103/PhysRevD.103.044031}}.

\bibitem{Rasouli:2022tmc}
S.~M.~M. Rasouli, S.~Jalalzadeh, P.~Moniz, {Noncompactified
  Kaluza\textendash{}Klein Gravity}, Universe 8~(8) (2022) 431.
\newblock \href {http://arxiv.org/abs/2208.11664} {\path{arXiv:2208.11664}},
  \href {http://dx.doi.org/10.3390/universe8080431}
  {\path{doi:10.3390/universe8080431}}.

\bibitem{Deffayet:2001pu}
C.~Deffayet, G.~R. Dvali, G.~Gabadadze, {Accelerated universe from gravity
  leaking to extra dimensions}, Phys. Rev. D 65 (2002) 044023.
\newblock \href {http://arxiv.org/abs/astro-ph/0105068}
  {\path{arXiv:astro-ph/0105068}}, \href
  {http://dx.doi.org/10.1103/PhysRevD.65.044023}
  {\path{doi:10.1103/PhysRevD.65.044023}}.

\bibitem{Deffayet:2002sp}
C.~Deffayet, S.~J. Landau, J.~Raux, M.~Zaldarriaga, P.~Astier, {Supernovae,
  CMB, and gravitational leakage into extra dimensions}, Phys. Rev. D 66 (2002)
  024019.
\newblock \href {http://arxiv.org/abs/astro-ph/0201164}
  {\path{arXiv:astro-ph/0201164}}, \href
  {http://dx.doi.org/10.1103/PhysRevD.66.024019}
  {\path{doi:10.1103/PhysRevD.66.024019}}.

\bibitem{Dvali:2000xg}
G.~R. Dvali, G.~Gabadadze, {Gravity on a brane in infinite volume extra space},
  Phys. Rev. D 63 (2001) 065007.
\newblock \href {http://arxiv.org/abs/hep-th/0008054}
  {\path{arXiv:hep-th/0008054}}, \href
  {http://dx.doi.org/10.1103/PhysRevD.63.065007}
  {\path{doi:10.1103/PhysRevD.63.065007}}.

\bibitem{Deffayet:2000uy}
C.~Deffayet, {Cosmology on a brane in Minkowski bulk}, Phys. Lett. B 502 (2001)
  199--208.
\newblock \href {http://arxiv.org/abs/hep-th/0010186}
  {\path{arXiv:hep-th/0010186}}, \href
  {http://dx.doi.org/10.1016/S0370-2693(01)00160-5}
  {\path{doi:10.1016/S0370-2693(01)00160-5}}.

\bibitem{Randall:1999vf}
L.~Randall, R.~Sundrum, {An Alternative to compactification}, Phys. Rev. Lett.
  83 (1999) 4690--4693.
\newblock \href {http://arxiv.org/abs/hep-th/9906064}
  {\path{arXiv:hep-th/9906064}}, \href
  {http://dx.doi.org/10.1103/PhysRevLett.83.4690}
  {\path{doi:10.1103/PhysRevLett.83.4690}}.

\bibitem{2004CQGra611M}
B.~{Mashhoon}, P.~S. {Wesson}, {Gauge-dependent cosmological 'constant'},
  Classical and Quantum Gravity 21~(14) (2004) 3611--3620.
\newblock \href {http://arxiv.org/abs/gr-qc/0401002}
  {\path{arXiv:gr-qc/0401002}}, \href
  {http://dx.doi.org/10.1088/0264-9381/21/14/020}
  {\path{doi:10.1088/0264-9381/21/14/020}}.

\bibitem{doi:10.1142/6029}
P.~S. Wesson,
  \href{https://www.worldscientific.com/doi/abs/10.1142/6029}{Five-Dimensional
  Physics: Classical and Quantum Consequences of Kaluza-Klein Cosmology}, World
  Scientific, 2006.
\newblock \href {http://dx.doi.org/10.1142/6029} {\path{doi:10.1142/6029}}.
\newline\urlprefix\url{https://www.worldscientific.com/doi/abs/10.1142/6029}

\bibitem{doi:10.1142/10871}
P.~S. Wesson, J.~M. Overduin,
  \href{https://www.worldscientific.com/doi/abs/10.1142/10871}{Principles of
  Space-Time-Matter}, World Scientific, 2018.
\newblock \href {http://dx.doi.org/10.1142/10871} {\path{doi:10.1142/10871}}.
\newline\urlprefix\url{https://www.worldscientific.com/doi/abs/10.1142/10871}

\bibitem{Jalalzadeh:2006mr}
S.~Jalalzadeh, {Non-integrability and Mach's principle in Induced Matter
  Theory}, Gen. Rel. Grav. 39 (2007) 387.
\newblock \href {http://arxiv.org/abs/gr-qc/0612090}
  {\path{arXiv:gr-qc/0612090}}, \href
  {http://dx.doi.org/10.1007/s10714-006-0391-1}
  {\path{doi:10.1007/s10714-006-0391-1}}.

\bibitem{Jalalzadeh:2008xu}
S.~Jalalzadeh, A.~M. Yazdani, {Variation of mass in primordial nucleosynthesis
  as a test of Induced Matter Brane Gravity}, Phys. Lett. B 664 (2008)
  229--234.
\newblock \href {http://arxiv.org/abs/0805.3017} {\path{arXiv:0805.3017}},
  \href {http://dx.doi.org/10.1016/j.physletb.2008.05.041}
  {\path{doi:10.1016/j.physletb.2008.05.041}}.

\bibitem{1986GReGr695M}
M.~D. {Maia}, {On Kaluza-Klein relativity.}, Gen. Relativ. Gravit. 18~(7)
  (1986) 695--699.
\newblock \href {http://dx.doi.org/10.1007/BF00768633}
  {\path{doi:10.1007/BF00768633}}.

\bibitem{1994PhRvD7233M}
M.~D. {Maia}, G.~S. {Silva}, {Geometrical constraints on the cosmological
  constant}, Phys. Rev. D 50~(12) (1994) 7233--7238.
\newblock \href {http://arxiv.org/abs/gr-qc/9401005}
  {\path{arXiv:gr-qc/9401005}}, \href
  {http://dx.doi.org/10.1103/PhysRevD.50.7233}
  {\path{doi:10.1103/PhysRevD.50.7233}}.

\bibitem{2002IJMPA341M}
M.~D. {Maia}, E.~M. {Monte}, J.~M.~F. {Maia}, {On Friedmann's Equation in
  Brane-Worlds}, Int. J. Mod. Phy. A 17~(29) (2002) 4341--4348.
\newblock \href {http://dx.doi.org/10.1142/S0217751X02013393}
  {\path{doi:10.1142/S0217751X02013393}}.

\bibitem{2002PhLA9M}
M.~D. {Maia}, E.~M. {Monte}, {Geometry of brane-worlds}, Phys. Lett. A
  297~(1-2) (2002) 9--19.
\newblock \href {http://arxiv.org/abs/hep-th/0110088}
  {\path{arXiv:hep-th/0110088}}, \href
  {http://dx.doi.org/10.1016/S0375-9601(02)00182-2}
  {\path{doi:10.1016/S0375-9601(02)00182-2}}.

\bibitem{2011GReGr2685M}
M.~D. Maia, A.~J.~S. Capistrano, J.~S. Alcaniz, E.~M. Monte, {The Deformable
  Universe}, Gen. Rel. Grav. 43 (2011) 2685--2700.
\newblock \href {http://arxiv.org/abs/1101.3951} {\path{arXiv:1101.3951}},
  \href {http://dx.doi.org/10.1007/s10714-011-1192-8}
  {\path{doi:10.1007/s10714-011-1192-8}}.

\bibitem{Jalalzadeh:2013wza}
S.~Jalalzadeh, T.~Rostami, {Covariant extrinsic gravity and the geometric
  origin of dark energy}, Int. J. Mod. Phys. D 24~(03) (2015) 1550027.
\newblock \href {http://arxiv.org/abs/1307.1913} {\path{arXiv:1307.1913}},
  \href {http://dx.doi.org/10.1142/S0218271815500273}
  {\path{doi:10.1142/S0218271815500273}}.

\bibitem{Rostami:2015ixa}
T.~Rostami, S.~Jalalzadeh, {Why the measured cosmological constant is small},
  Phys. Dark Univ. 9-10 (2015) 31--36.
\newblock \href {http://arxiv.org/abs/1510.02068} {\path{arXiv:1510.02068}},
  \href {http://dx.doi.org/10.1016/j.dark.2015.10.001}
  {\path{doi:10.1016/j.dark.2015.10.001}}.

\bibitem{Heydari-Fard:2006klr}
M.~Heydari-Fard, M.~Shirazi, S.~Jalalzadeh, H.~R. Sepangi, {Accelerating
  universe in brane gravity with a confining potential}, Phys. Lett. B 640
  (2006) 1--6.
\newblock \href {http://arxiv.org/abs/gr-qc/0607067}
  {\path{arXiv:gr-qc/0607067}}, \href
  {http://dx.doi.org/10.1016/j.physletb.2006.07.020}
  {\path{doi:10.1016/j.physletb.2006.07.020}}.

\bibitem{Davis:2007na}
T.~M. Davis, et~al., {Scrutinizing Exotic Cosmological Models Using ESSENCE
  Supernova Data Combined with Other Cosmological Probes}, Astrophys. J. 666
  (2007) 716--725.
\newblock \href {http://arxiv.org/abs/astro-ph/0701510}
  {\path{arXiv:astro-ph/0701510}}, \href {http://dx.doi.org/10.1086/519988}
  {\path{doi:10.1086/519988}}.

\bibitem{Rubin:2008wq}
D.~Rubin, et~al., {Looking Beyond Lambda with the Union Supernova Compilation},
  Astrophys. J. 695 (2009) 391--403.
\newblock \href {http://arxiv.org/abs/0807.1108} {\path{arXiv:0807.1108}},
  \href {http://dx.doi.org/10.1088/0004-637X/695/1/391}
  {\path{doi:10.1088/0004-637X/695/1/391}}.

\bibitem{Song:2006jk}
Y.-S. Song, I.~Sawicki, W.~Hu, {Large-Scale Tests of the DGP Model}, Phys. Rev.
  D 75 (2007) 064003.
\newblock \href {http://arxiv.org/abs/astro-ph/0606286}
  {\path{arXiv:astro-ph/0606286}}, \href
  {http://dx.doi.org/10.1103/PhysRevD.75.064003}
  {\path{doi:10.1103/PhysRevD.75.064003}}.

\bibitem{Schmidt:2009sg}
F.~Schmidt, {Self-Consistent Cosmological Simulations of DGP Braneworld
  Gravity}, Phys. Rev. D 80 (2009) 043001.
\newblock \href {http://arxiv.org/abs/0905.0858} {\path{arXiv:0905.0858}},
  \href {http://dx.doi.org/10.1103/PhysRevD.80.043001}
  {\path{doi:10.1103/PhysRevD.80.043001}}.

\bibitem{Fang:2008kc}
W.~Fang, S.~Wang, W.~Hu, Z.~Haiman, L.~Hui, M.~May, {Challenges to the DGP
  Model from Horizon-Scale Growth and Geometry}, Phys. Rev. D 78 (2008) 103509.
\newblock \href {http://arxiv.org/abs/0808.2208} {\path{arXiv:0808.2208}},
  \href {http://dx.doi.org/10.1103/PhysRevD.78.103509}
  {\path{doi:10.1103/PhysRevD.78.103509}}.

\bibitem{Guo:2006ce}
Z.-K. Guo, Z.-H. Zhu, J.~S. Alcaniz, Y.-Z. Zhang, {Constraints on the dgp model
  from recent supernova observations and baryon acoustic oscillations},
  Astrophys. J. 646 (2006) 1--7.
\newblock \href {http://arxiv.org/abs/astro-ph/0603632}
  {\path{arXiv:astro-ph/0603632}}, \href {http://dx.doi.org/10.1086/504831}
  {\path{doi:10.1086/504831}}.

\bibitem{Fairbairn:2005ue}
M.~Fairbairn, A.~Goobar, {Supernova limits on brane world cosmology}, Phys.
  Lett. B 642 (2006) 432--435.
\newblock \href {http://arxiv.org/abs/astro-ph/0511029}
  {\path{arXiv:astro-ph/0511029}}, \href
  {http://dx.doi.org/10.1016/j.physletb.2006.07.048}
  {\path{doi:10.1016/j.physletb.2006.07.048}}.

\bibitem{Maartens:2006yt}
R.~Maartens, E.~Majerotto, {Observational constraints on self-accelerating
  cosmology}, Phys. Rev. D 74 (2006) 023004.
\newblock \href {http://arxiv.org/abs/astro-ph/0603353}
  {\path{arXiv:astro-ph/0603353}}, \href
  {http://dx.doi.org/10.1103/PhysRevD.74.023004}
  {\path{doi:10.1103/PhysRevD.74.023004}}.

\bibitem{Xia:2009gb}
J.-Q. Xia, {Constraining DGP Gravity from Observational Data}, Phys. Rev. D 79
  (2009) 103527.
\newblock \href {http://arxiv.org/abs/0907.4860} {\path{arXiv:0907.4860}},
  \href {http://dx.doi.org/10.1103/PhysRevD.79.103527}
  {\path{doi:10.1103/PhysRevD.79.103527}}.

\bibitem{Dubovsky:2002jm}
S.~L. Dubovsky, V.~A. Rubakov, {Brane induced gravity in more than one extra
  dimensions: Violation of equivalence principle and ghost}, Phys. Rev. D 67
  (2003) 104014.
\newblock \href {http://arxiv.org/abs/hep-th/0212222}
  {\path{arXiv:hep-th/0212222}}, \href
  {http://dx.doi.org/10.1103/PhysRevD.67.104014}
  {\path{doi:10.1103/PhysRevD.67.104014}}.

\bibitem{deRham:2007xp}
C.~de~Rham, G.~Dvali, S.~Hofmann, J.~Khoury, O.~Pujolas, M.~Redi, A.~J. Tolley,
  {Cascading gravity: Extending the Dvali-Gabadadze-Porrati model to higher
  dimension}, Phys. Rev. Lett. 100 (2008) 251603.
\newblock \href {http://arxiv.org/abs/0711.2072} {\path{arXiv:0711.2072}},
  \href {http://dx.doi.org/10.1103/PhysRevLett.100.251603}
  {\path{doi:10.1103/PhysRevLett.100.251603}}.

\bibitem{Nicolis:2008in}
A.~Nicolis, R.~Rattazzi, E.~Trincherini, {The Galileon as a local modification
  of gravity}, Phys. Rev. D 79 (2009) 064036.
\newblock \href {http://arxiv.org/abs/0811.2197} {\path{arXiv:0811.2197}},
  \href {http://dx.doi.org/10.1103/PhysRevD.79.064036}
  {\path{doi:10.1103/PhysRevD.79.064036}}.

\bibitem{Zhang:2004in}
H.-s. Zhang, R.-G. Cai, {Inflation on Dvali-Gabadadze-Porrati brane}, JCAP 08
  (2004) 017.
\newblock \href {http://arxiv.org/abs/hep-th/0403234}
  {\path{arXiv:hep-th/0403234}}, \href
  {http://dx.doi.org/10.1088/1475-7516/2004/08/017}
  {\path{doi:10.1088/1475-7516/2004/08/017}}.

\bibitem{Garcia-Aspeitia:2020snv}
M.~A. Garc\'\i{}a-Aspeitia, C.~Escamilla-Rivera, {Gravitational waves in
  braneworlds after multi-messenger events}, Eur. Phys. J. C 80~(4) (2020) 316.
\newblock \href {http://arxiv.org/abs/2001.08745} {\path{arXiv:2001.08745}},
  \href {http://dx.doi.org/10.1140/epjc/s10052-020-7895-9}
  {\path{doi:10.1140/epjc/s10052-020-7895-9}}.

\bibitem{2013ApJS19H}
H.~G. et~al, {Nine-year Wilkinson Microwave Anisotropy Probe (WMAP)
  Observations: Cosmological Parameter Results}, APJS 208~(2) (2013) 19.
\newblock \href {http://arxiv.org/abs/1212.5226} {\path{arXiv:1212.5226}},
  \href {http://dx.doi.org/10.1088/0067-0049/208/2/19}
  {\path{doi:10.1088/0067-0049/208/2/19}}.

\bibitem{Maia:1984nv}
M.~D. Maia, {GEOMETRY OF KALUZA-KLEIN THEORY. I. BASIC SETTING}, Phys. Rev. D
  31 (1985) 262--267.
\newblock \href {http://dx.doi.org/10.1103/PhysRevD.31.262}
  {\path{doi:10.1103/PhysRevD.31.262}}.

\bibitem{Maia:1983zh}
M.~D. Maia, W.~Mecklenburg, {Aspects of high dimensional theories in embedding
  spaces}, J. Math. Phys. 25 (1984) 3047.
\newblock \href {http://dx.doi.org/10.1063/1.526020}
  {\path{doi:10.1063/1.526020}}.

\bibitem{Eisenhart}
L.~P. Eisenhart, Riemannian Geometry, Vol.~19 of Princeton Landmarks in
  Mathematics and Physics, Princeton University Press, 1966.

\bibitem{Bueno:2022log}
P.~Bueno, R.~Emparan, Q.~Llorens, {Higher-curvature gravities from braneworlds
  and the holographic c-theorem}, Phys. Rev. D 106~(4) (2022) 044012.
\newblock \href {http://arxiv.org/abs/2204.13421} {\path{arXiv:2204.13421}},
  \href {http://dx.doi.org/10.1103/PhysRevD.106.044012}
  {\path{doi:10.1103/PhysRevD.106.044012}}.

\bibitem{twist}
E.~M. Monte, M.~D. Maia, {The twisting connection of space-time}, J. Math.
  Phys. 37 (1996) 1972--1981.
\newblock \href {http://dx.doi.org/10.1063/1.531488}
  {\path{doi:10.1063/1.531488}}.

\bibitem{Yang}
M.~D. Maia, {On the Integrability Conditions for Extended Objects}, Class.
  Quant. Grav. 6 (1989) 173--183.
\newblock \href {http://dx.doi.org/10.1088/0264-9381/6/2/011}
  {\path{doi:10.1088/0264-9381/6/2/011}}.

\bibitem{Shahram}
S.~Jalalzadeh, B.~Vakili, H.~R. Sepangi, {On extra forces from large extra
  dimensions}, Phys. Scripta 76 (2007) 122--126.
\newblock \href {http://arxiv.org/abs/gr-qc/0409070}
  {\path{arXiv:gr-qc/0409070}}, \href
  {http://dx.doi.org/10.1088/0031-8949/76/2/002}
  {\path{doi:10.1088/0031-8949/76/2/002}}.

\bibitem{Jalalzadeh:2004uv}
S.~Jalalzadeh, H.~R. Sepangi, {Classical and quantum dynamics of confined test
  particles in brane gravity}, Class. Quant. Grav. 22 (2005) 2035--2048.
\newblock \href {http://arxiv.org/abs/gr-qc/0408004}
  {\path{arXiv:gr-qc/0408004}}, \href
  {http://dx.doi.org/10.1088/0264-9381/22/11/008}
  {\path{doi:10.1088/0264-9381/22/11/008}}.

\bibitem{Love}
D.~Bailin, A.~Love, {KALUZA-KLEIN THEORIES}, Rept. Prog. Phys. 50 (1987)
  1087--1170.
\newblock \href {http://dx.doi.org/10.1088/0034-4885/50/9/001}
  {\path{doi:10.1088/0034-4885/50/9/001}}.

\bibitem{Maia:2004fq}
M.~D. Maia, E.~M. Monte, J.~M.~F. Maia, J.~S. Alcaniz, {On the geometry of dark
  energy}, Class. Quant. Grav. 22 (2005) 1623--1636.
\newblock \href {http://arxiv.org/abs/astro-ph/0403072}
  {\path{arXiv:astro-ph/0403072}}, \href
  {http://dx.doi.org/10.1088/0264-9381/22/9/010}
  {\path{doi:10.1088/0264-9381/22/9/010}}.

\bibitem{Williams:2003wu}
J.~G. Williams, S.~G. Turyshev, T.~W. Murphy, Jr., {Improving LLR tests of
  gravitational theory}, Int. J. Mod. Phys. D 13 (2004) 567--582.
\newblock \href {http://arxiv.org/abs/gr-qc/0311021}
  {\path{arXiv:gr-qc/0311021}}, \href
  {http://dx.doi.org/10.1142/S0218271804004682}
  {\path{doi:10.1142/S0218271804004682}}.

\bibitem{Merkowitz:2010kka}
S.~M. Merkowitz, {Tests of Gravity Using Lunar Laser Ranging}, Living Rev. Rel.
  13 (2010) 7.
\newblock \href {http://dx.doi.org/10.12942/lrr-2010-7}
  {\path{doi:10.12942/lrr-2010-7}}.

\bibitem{Gaztanaga:2001fh}
E.~Gaztanaga, E.~Garcia-Berro, J.~Isern, E.~Bravo, I.~Dominguez, {Bounds on the
  possible evolution of the gravitational constant from cosmological type Ia
  supernovae}, Phys. Rev. D 65 (2002) 023506.
\newblock \href {http://arxiv.org/abs/astro-ph/0109299}
  {\path{arXiv:astro-ph/0109299}}, \href
  {http://dx.doi.org/10.1103/PhysRevD.65.023506}
  {\path{doi:10.1103/PhysRevD.65.023506}}.

\bibitem{1998ApJ871G}
D.~B. {Guenther}, L.~M. {Krauss}, P.~{Demarque}, {Testing the Constancy of the
  Gravitational Constant Using Helioseismology}, Astrophys. J. 498~(2) (1998)
  871--876.
\newblock \href {http://dx.doi.org/10.1086/305567} {\path{doi:10.1086/305567}}.

\bibitem{Damour:1988zz}
T.~Damour, G.~W. Gibbons, J.~H. Taylor, {Limits on the Variability of G Using
  Binary-Pulsar Data}, Phys. Rev. Lett. 61 (1988) 1151--1154.
\newblock \href {http://dx.doi.org/10.1103/PhysRevLett.61.1151}
  {\path{doi:10.1103/PhysRevLett.61.1151}}.

\bibitem{Doroud:2009zza}
N.~Doroud, S.~M.~M. Rasouli, S.~Jalalzadeh, {A class of cosmological solutions
  in induced matter theory with conformally flat bulk space}, Gen. Rel. Grav.
  41 (2009) 2637--2656.
\newblock \href {http://dx.doi.org/10.1007/s10714-009-0793-y}
  {\path{doi:10.1007/s10714-009-0793-y}}.

\bibitem{Moyassari:2007sv}
P.~Moyassari, S.~Jalalzadeh, {Semiclassical corrections to the Einstein
  equation and induced matter theory}, Gen. Rel. Grav. 39 (2007) 1467--1476.
\newblock \href {http://arxiv.org/abs/0705.2289} {\path{arXiv:0705.2289}},
  \href {http://dx.doi.org/10.1007/s10714-007-0466-7}
  {\path{doi:10.1007/s10714-007-0466-7}}.

\bibitem{Jalalzadeh:2006nh}
S.~Jalalzadeh, B.~Vakili, F.~Ahmadi, H.~R. Sepangi, {Stabilization of test
  particles in induced matter Kaluza-Klein theory}, Class. Quant. Grav. 23
  (2006) 6015--6030.
\newblock \href {http://arxiv.org/abs/gr-qc/0608071}
  {\path{arXiv:gr-qc/0608071}}, \href
  {http://dx.doi.org/10.1088/0264-9381/23/20/021}
  {\path{doi:10.1088/0264-9381/23/20/021}}.

\bibitem{Shiromizu:1999wj}
T.~Shiromizu, K.-i. Maeda, M.~Sasaki, {The Einstein equation on the 3-brane
  world}, Phys. Rev. D 62 (2000) 024012.
\newblock \href {http://arxiv.org/abs/gr-qc/9910076}
  {\path{arXiv:gr-qc/9910076}}, \href
  {http://dx.doi.org/10.1103/PhysRevD.62.024012}
  {\path{doi:10.1103/PhysRevD.62.024012}}.

\bibitem{Faraoni:2021opj}
V.~Faraoni, S.~Jose, S.~Dussault, {Multi-fluid cosmology in Einstein gravity:
  analytical solutions}, Gen. Rel. Grav. 53~(12) (2021) 109.
\newblock \href {http://arxiv.org/abs/2107.12488} {\path{arXiv:2107.12488}},
  \href {http://dx.doi.org/10.1007/s10714-021-02879-z}
  {\path{doi:10.1007/s10714-021-02879-z}}.

\bibitem{1986GReGr79N}
B.~K. {Nayak}, G.~B. {Bhuyan}, {Bianchi type-V perfect fluid models.}, Gen.
  Rel. Grav. 18~(1) (1986) 79--91.
\newblock \href {http://dx.doi.org/10.1007/BF00843752}
  {\path{doi:10.1007/BF00843752}}.

\bibitem{2019Ph2A}
{\"O}.~{Akarsu}, S.~{Kumar}, S.~{Sharma}, L.~{Tedesco}, {Constraints on a
  Bianchi type I spacetime extension of the standard {\ensuremath{\Lambda}} CDM
  model}, Phys. Rev. D 100~(2) (2019) 023532.
\newblock \href {http://arxiv.org/abs/1905.06949} {\path{arXiv:1905.06949}},
  \href {http://dx.doi.org/10.1103/PhysRevD.100.023532}
  {\path{doi:10.1103/PhysRevD.100.023532}}.

\bibitem{Riess:2019cxk}
A.~G. Riess, S.~Casertano, W.~Yuan, L.~M. Macri, D.~Scolnic, {Large Magellanic
  Cloud Cepheid Standards Provide a 1\% Foundation for the Determination of the
  Hubble Constant and Stronger Evidence for Physics beyond $\Lambda$CDM},
  Astrophys. J. 876~(1) (2019) 85.
\newblock \href {http://arxiv.org/abs/1903.07603} {\path{arXiv:1903.07603}},
  \href {http://dx.doi.org/10.3847/1538-4357/ab1422}
  {\path{doi:10.3847/1538-4357/ab1422}}.

\bibitem{Planck:2018vyg}
N.~Aghanim, et~al., {Planck 2018 results. VI. Cosmological parameters}, Astron.
  Astrophys. 641 (2020) A6, [Erratum: Astron.Astrophys. 652, C4 (2021)].
\newblock \href {http://arxiv.org/abs/1807.06209} {\path{arXiv:1807.06209}},
  \href {http://dx.doi.org/10.1051/0004-6361/201833910}
  {\path{doi:10.1051/0004-6361/201833910}}.

\bibitem{Lovelace:1971fa}
C.~Lovelace, {Pomeron form-factors and dual Regge cuts}, Phys. Lett. B 34
  (1971) 500--506.
\newblock \href {http://dx.doi.org/10.1016/0370-2693(71)90665-4}
  {\path{doi:10.1016/0370-2693(71)90665-4}}.

\bibitem{Cvetic:2011vz}
M.~Cvetic, J.~Halverson, {TASI Lectures: Particle Physics from Perturbative and
  Non-perturbative Effects in D-braneworlds}, in: {Theoretical Advanced Study
  Institute in Elementary Particle Physics}: {String theory and its
  Applications: From meV to the Planck Scale}, 2011, pp. 245--292.
\newblock \href {http://arxiv.org/abs/1101.2907} {\path{arXiv:1101.2907}},
  \href {http://dx.doi.org/10.1142/9789814350525-0005}
  {\path{doi:10.1142/9789814350525-0005}}.

\bibitem{Lineweaver:2003ie}
C.~H. Lineweaver, {Inflation and the cosmic microwave background}, in: {16th
  Canberra International Physics Summer School: The New Cosmology}, 2003.
\newblock \href {http://arxiv.org/abs/astro-ph/0305179}
  {\path{arXiv:astro-ph/0305179}}.

\bibitem{Zeldovich:1967gd}
Y.~B. Zeldovich, {Cosmological Constant and Elementary Particles}, JETP Lett. 6
  (1967) 316.

\bibitem{Weinberg:1972kfs}
S.~Weinberg, {Gravitation and Cosmology}: {Principles and Applications of the
  General Theory of Relativity}, John Wiley and Sons, New York, 1972.

\bibitem{Wesson:2003qn}
P.~S. Wesson, {Is mass quantized?}, Mod. Phys. Lett. A 19 (2004) 1995--2000.
\newblock \href {http://arxiv.org/abs/gr-qc/0309100}
  {\path{arXiv:gr-qc/0309100}}, \href
  {http://dx.doi.org/10.1142/S0217732304015270}
  {\path{doi:10.1142/S0217732304015270}}.

\bibitem{Yazdani:2011ben}
A.~M. Yazdani, K.~Atazadeh, S.~Jalalzadeh, {Localization of gravity in brane
  world with arbitrary extra dimensions}, Int. J. Theor. Phys. 50 (2011)
  888--898.
\newblock \href {http://arxiv.org/abs/1203.3762} {\path{arXiv:1203.3762}},
  \href {http://dx.doi.org/10.1007/s10773-010-0630-9}
  {\path{doi:10.1007/s10773-010-0630-9}}.

\bibitem{Jalalzadeh:2023liy}
S.~Jalalzadeh, {Intrinsic quantum dynamics of particles in brane gravity},
  Annals Phys. 452 (2023) 169291.
\newblock \href {http://arxiv.org/abs/2303.11104} {\path{arXiv:2303.11104}},
  \href {http://dx.doi.org/10.1016/j.aop.2023.169291}
  {\path{doi:10.1016/j.aop.2023.169291}}.

\bibitem{Frye:2012jj}
C.~R. Frye, C.~J. Efthimiou,
  \href{https://www.worldscientific.com/doi/abs/10.1142/9134}{{Spherical
  Harmonics in p Dimensions}}, {World Scientific}, {2014}.
\newblock \href {http://arxiv.org/abs/{1205.3548}} {\path{arXiv:{1205.3548}}},
  \href {http://dx.doi.org/{10.1142/9134}} {\path{doi:{10.1142/9134}}}.
\newline\urlprefix\url{https://www.worldscientific.com/doi/abs/10.1142/9134}

\bibitem{Dvali:2002vf}
G.~Dvali, A.~Gruzinov, M.~Zaldarriaga, {The Accelerated universe and the moon},
  Phys. Rev. D 68 (2003) 024012.
\newblock \href {http://arxiv.org/abs/hep-ph/0212069}
  {\path{arXiv:hep-ph/0212069}}, \href
  {http://dx.doi.org/10.1103/PhysRevD.68.024012}
  {\path{doi:10.1103/PhysRevD.68.024012}}.

\bibitem{Lue:2002sw}
A.~Lue, G.~Starkman, {Gravitational leakage into extra dimensions: Probing dark
  energy using local gravity}, Phys. Rev. D 67 (2003) 064002.
\newblock \href {http://arxiv.org/abs/astro-ph/0212083}
  {\path{arXiv:astro-ph/0212083}}, \href
  {http://dx.doi.org/10.1103/PhysRevD.67.064002}
  {\path{doi:10.1103/PhysRevD.67.064002}}.

\bibitem{Choudhury:2002pu}
S.~R. Choudhury, G.~C. Joshi, S.~Mahajan, B.~H.~J. McKellar, {Probing large
  distance higher dimensional gravity from lensing data}, Astropart. Phys. 21
  (2004) 559--563.
\newblock \href {http://arxiv.org/abs/hep-ph/0204161}
  {\path{arXiv:hep-ph/0204161}}, \href
  {http://dx.doi.org/10.1016/j.astropartphys.2004.04.001}
  {\path{doi:10.1016/j.astropartphys.2004.04.001}}.

\bibitem{Gruzinov:2001hp}
A.~Gruzinov, {On the graviton mass}, New Astron. 10 (2005) 311--314.
\newblock \href {http://arxiv.org/abs/astro-ph/0112246}
  {\path{arXiv:astro-ph/0112246}}, \href
  {http://dx.doi.org/10.1016/j.newast.2004.12.001}
  {\path{doi:10.1016/j.newast.2004.12.001}}.

\bibitem{Gazeau:2006uy}
J.-P. Gazeau, M.~Novello, {The Nature of Lambda and the mass of the graviton: A
  Critical view}, Int. J. Mod. Phys. A 26 (2011) 3697--3720.
\newblock \href {http://arxiv.org/abs/gr-qc/0610054}
  {\path{arXiv:gr-qc/0610054}}, \href
  {http://dx.doi.org/10.1142/S0217751X11054176}
  {\path{doi:10.1142/S0217751X11054176}}.

\bibitem{Bousso:2000nf}
R.~Bousso, {Positive vacuum energy and the N bound}, JHEP 11 (2000) 038.
\newblock \href {http://arxiv.org/abs/hep-th/0010252}
  {\path{arXiv:hep-th/0010252}}, \href
  {http://dx.doi.org/10.1088/1126-6708/2000/11/038}
  {\path{doi:10.1088/1126-6708/2000/11/038}}.

\bibitem{Mongan:2006ay}
T.~R. Mongan, {Holography and non-locality in a closed vacuum-dominated
  universe}, Int. J. Theor. Phys. 46 (2007) 399--404.
\newblock \href {http://arxiv.org/abs/gr-qc/0609128}
  {\path{arXiv:gr-qc/0609128}}, \href
  {http://dx.doi.org/10.1007/s10773-006-9241-x}
  {\path{doi:10.1007/s10773-006-9241-x}}.

\bibitem{Jalalzadeh:2017jdo}
S.~Jalalzadeh, A.~J.~S. Capistrano, P.~V. Moniz, {Quantum deformation of
  quantum cosmology: A framework to discuss the cosmological constant problem},
  Phys. Dark Univ. 18 (2017) 55--66.
\newblock \href {http://arxiv.org/abs/1709.09923} {\path{arXiv:1709.09923}},
  \href {http://dx.doi.org/10.1016/j.dark.2017.09.011}
  {\path{doi:10.1016/j.dark.2017.09.011}}.

\bibitem{201825E}
C.~A. {Egan}, C.~H. {Lineweaver}, {A Larger Estimate of the Entropy of the
  Universe}, Astrophys. J. 710~(2) (2010) 1825--1834.
\newblock \href {http://arxiv.org/abs/0909.3983} {\path{arXiv:0909.3983}},
  \href {http://dx.doi.org/10.1088/0004-637X/710/2/1825}
  {\path{doi:10.1088/0004-637X/710/2/1825}}.

\bibitem{Carroll:2000fy}
S.~M. Carroll, {The Cosmological constant}, Living Rev. Rel. 4 (2001) 1.
\newblock \href {http://arxiv.org/abs/astro-ph/0004075}
  {\path{arXiv:astro-ph/0004075}}, \href
  {http://dx.doi.org/10.12942/lrr-2001-1} {\path{doi:10.12942/lrr-2001-1}}.

\bibitem{Weinberg:1988cp}
S.~Weinberg, {The Cosmological Constant Problem}, Rev. Mod. Phys. 61 (1989)
  1--23.
\newblock \href {http://dx.doi.org/10.1103/RevModPhys.61.1}
  {\path{doi:10.1103/RevModPhys.61.1}}.

\end{thebibliography}

\end{document}